\documentclass{article}

\usepackage{arxiv}

\usepackage[utf8]{inputenc} % allow utf-8 input
\usepackage[T1]{fontenc}    % use 8-bit T1 fonts
\usepackage{hyperref}       % hyperlinks
\usepackage{url}            % simple URL typesetting
\usepackage{booktabs}       % professional-quality tables
\usepackage{amsfonts}       % blackboard math symbols
\usepackage{nicefrac}       % compact symbols for 1/2, etc.
\usepackage{microtype}      % microtypography
\usepackage{lipsum}

\usepackage{breqn}
\usepackage{amssymb}
\usepackage{amsthm}
\usepackage{graphicx}
\usepackage{caption}
\usepackage{color}
\usepackage{caption}
\usepackage{subcaption}
\usepackage{color}
\usepackage{wrapfig}

\newtheorem{definition}{Definition}

\newcommand{\red}[1]{\textcolor{black}{#1}}

\usepackage[inline]{enumitem}

\title{Network connectivity under a probabilistic node failure model}

\author{
  Lucia~Cavallaro\\
  Data Science Research Centre\\
  University of Derby (UK)\\
  \texttt{l.cavallaro@derby.ac.uk} \\
   \And
  Stefania~Costantini \\
  Department of Computer Science\\
  University of L'Aquila (Italy)\\
  \texttt{stefania.costantini@univaq.it} \\
   \AND
   Pasquale~De~Meo \\
   Department of Ancient and Modern Civilizations \\
   University of Messina (Italy) \\
   \texttt{ pdemeo@unime.it} \\
   \And
   Antonio~Liotta \\
   Faculty of Computer Science \\
   Free University of Bozen-Bolzano (Italy) \\
   \texttt{liotta.antonio@gmail.com} \\
   \And
   Giovanni~Stilo \\
   Department of Computer Science\\
   University of L'Aquila (Italy)\\
   \texttt{giovanni.stilo@univaq.it} \\
}

\begin{document}
\maketitle

\begin{abstract}
Centrality metrics have been widely applied to identify the nodes in a graph whose removal is effective in decomposing the graph into smaller sub-components. The node--removal process is generally used to test network robustness against failures. Most of the available studies assume that the node removal task is always successful. Yet, we argue that this assumption is unrealistic. Indeed, the removal process should take into account also the strength of the targeted node itself, to simulate the failure scenarios in a more effective and realistic fashion. Unlike previous literature, herein a {\em probabilistic node failure model} is proposed, in which nodes may fail with a particular probability, considering two variants, namely: {\em Uniform} (in which the nodes survival-to-failure probability is fixed) and {\em Best Connected} (BC) (where the nodes survival probability is proportional to their degree). To evaluate our method, we consider five popular centrality metrics carrying out an experimental, comparative analysis to evaluate them in terms of {\em effectiveness} and {\em coverage}, on four real-world graphs. By effectiveness and coverage we mean the ability of selecting nodes whose removal decreases graph connectivity the most. Specifically, the graph spectral radius reduction works as a proxy indicator of effectiveness, and the reduction of the largest connected component (LCC) size is a parameter to assess coverage. The metric that caused the biggest drop has been then compared with the Benchmark analysis (i.e, the non-probabilistic degree centrality node removal process) to compare the two approaches. The main finding has been that significant differences emerged through this comparison with a deviation range that varies from 2\% up to 80\% regardless of the dataset used that highlight the existence of a gap between the common practice with a more realistic approach.
\end{abstract}

% keywords can be removed
\keywords{Node Centrality in Networks \and Spectral Radius \and Largest Connected Component}

\section{Introduction}
Node removal process is an important strategy in Network Science to deal with complex networks because it allows to test the network strength by stressing it to verify its resilience. On the other hand, another common application is to disconnect sub-graphs with the aim to prune the network still maintaining its vital connections in order to streamline the overall graph.

A common strategy used to detect which nodes should be removed is by ranking them accordingly to centrality metrics (Sect.~\ref{sec:background}), such as the {\em degree}\cite{Newman10}, the {\em h-index}~\cite{hirsch2005hindex,korn2009lobby}, the {\em coreness centrality}~\cite{bae2014coreness,kitsak2010identification}, the {\em Eigenvector centrality}~\cite{bonacich1987power}, and the {\em Katz centrality}~\cite{katz1953new}. Indeed, those metrics are a widespread tool in Network Science to identify important elements ({\em nodes} and {\em edges}) in real-life networked systems.\footnote{Other popular centrality metrics are {\em Betweenness centrality} and {\em Closeness centrality}; yet their computation requires to calculate all-pairs shortest paths in a graph, which is prohibitive even in graphs of modest size. Thus, in this paper Betweenness and Closeness centrality have not been considered.}

Most of the previous studies (Sect.~\ref{sec:related-literature}) assume that any attempt at removing a node is {\em always successful}, i.e., that an arbitrary node will always fail if attacked~\cite{Newman10,lu2016vital,albert2004structural,albert2000error,MeoMRSV18}. 

We argue that this assumption is not always realistic. Indeed, if the goal is to analyse an engineer network, such as an energy network, it is intuitive to expect dealing with a more resilient network to external attacks (e.g. malicious hackers that want to disrupt the network) or failures (e.g., hardware failure) rather than other kinds of networks (such as social networks). 

Thus, moving from a theoretical approach on graphs to application domains on real networks may affect the effectiveness of the always successful nodes removal approach.

It leads to a main (and yet unexplored) technical challenge that consists of estimating the damages on the connectivity of a graph $G$, under the assumption that nodes may resist to failures. Nodes, in fact, may resist to failures (or attacks) with a certain resilience percentage that may be very high (or low) accordingly to the node's role within the network analysed. Those scenarios differ widely from the classical removal strategy (herein used as benchmark).

Thus, this paper introduces a {\em probabilistic node failure model} $\mathcal{M}$, which associates each node with its probability to survive a failure (Sect.~\ref{sec:network-reliability}). When the node survival probability is zero (herein addressed as ``Benchmark'' analysis), the model coincides with other models already introduced in the literature~\cite{albert2000error,albert2004structural}. 

We consider two {\em variants} of $\mathcal{M}$, namely {\em Uniform}, in which the probability that any node survives a failure is equal to a fixed value $p$, and {\em Best Connected} (BC), in which the probability that a node survives a failure is proportional to its degree. 

To quantify the loss of connectivity in $G$ under both the Uniform and BC models have been used, namely the {\em spectral radius} $\lambda_1$, and the {\em largest connected component} (LCC) size $c$. 

The spectral radius is defined as the adjacency matrix\footnote{The adjacency matrix $\mathbf{A}$ of a graph $G$ is defined as $\mathbf{A}_{ij} = 1$ if nodes $i$ and $j$ are connected, 0 otherwise.} largest eigenvalue of $G$, and governs a broad range of spreading processes in $G$, such as the diffusion of an infection~\cite{chen2016eigen,wang2003epidemic,ganesh2005effect,prakash2012threshold}, malware propagation~\cite{berger2005spread,kleinberg2007computing}, or the dissemination of fake news in Online Social Networks (OSNs)~\cite{jiang2017identifying,amoruso2017contrasting,lazer2018science}. The LCC size is defined as the number of nodes in the largest connected subgraph of $G$, and is widely used to quantify the resilience of a natural or artificial system described by the graph $G$~\cite{albert2000error}.

With those assumptions in mind, we finally simulated a {\em node removal process} in which nodes may fail according to the Uniform or the BC model. Failed nodes are sorted by the degree, h-index, coreness, Eigenvector, and Katz centrality scores. For brevity, $\phi$ herein denotes any of the centrality metrics above.
Next, the top $\lceil \tau \times \vert N \vert \rceil$ nodes from the node ranking generated by $\phi$ are deleted, being $N$ the set of nodes of $G$ and $\tau$ a fixed threshold in $[0, 1]$. This process generates a graph $\tilde{G}$ with spectral radius $\tilde{\lambda}_1$ and the LCC size $\tilde{c}$.

To verify the impact of our proposed models, $\lambda_1$ and $c$ translate in two metrics namely {\em effectiveness} and {\em coverage}, respectively. The first one is defined by $\rho(\tau, \phi)$ of $\phi$ as the ratio of $\tilde{\lambda_1}$ to $\lambda_1$; analogously, the {\em coverage} $\gamma(\tau, \phi)$ is defined as the ratio of $\tilde c$ to $c$. By construction, $\tilde{\lambda}_1 \leq \lambda_1$ and $\tilde{c} \leq c$, which implies that both $\rho(\tau, \phi)$ and $\gamma(\tau, \phi)$ always range between $0$ and $1$. The smaller the magnitude of $\rho(\tau, \phi)$ and $\gamma(\tau, \phi)$, the bigger the damage observed in the connectivity of $G$ upon node removal.

Four large real-life graphs are considered in order to evaluate both the effectiveness and the coverage of $\phi$, as function of $\tau$. These datasets include: \textsc{US-POWER\_GRID} (describing the power grid in US Western states), \textsc{AstroPh} (describing co-authorship between scientists in the astrophysics domain), \textsc{BrightKite} (a location-based social networking Web site), and \textsc{Flickr} (a graph whereby the nodes represent Flickr photos and an edge indicates that two photos share some tags). 

Furthermore, the centrality metrics above have been computed via {\em NetShield}~\cite{chen2016node}, a state-of-the-art approximation algorithm, which takes an integer $k$ as input and seeks at discovering the set of $k$ nodes which, if deleted from $G$, produces the biggest drop in $\lambda_1$.

The main outcomes of the study are as follows (Sect.~\ref{sec:experiments}): 
\begin{enumerate*}[label=(\roman*)] 
\item the degree centrality metric, on average, generates the biggest drop in both $\lambda_1$ and $c$. This is why we also compared the results obtained with the always successful node removal process based on the same metric (i.e., the ``Benchmark''). In addition, the degree centrality is a viable alternative to the NetShield algorithm to quickly discover group of nodes whose removal yields a relevant drop in the spectral radius.
\item The BC model reports a large drop in $\lambda_1$ even when only a small fraction of nodes actually fails. To achieve a comparable reduction in $\lambda_1$ in the Uniform model should, on average, be deleted more than $30\%$ of nodes.
\item In graphs deriving from human interactions and collaborations (namely \textsc{BrightKite}, \textsc{Flickr} and \textsc{AstroPh}) a bigger drop in $\lambda_1$ rather than in \textsc{US-POWER\_GRID} has been noticed.
\item The degree, h-index, and coreness display the same behaviour in social networks (such as \textsc{BrightKite}) and in collaborative networks (such as \textsc{AstroPh}), thus confirming insights provided in~\cite{lu2015hindex}.
\item The degree and the coreness yield the fastest decrease in coverage even if, in some datasets, such a decrease is only marginally smaller than that observed when the other centrality metrics discussed in this paper are applied.
\end{enumerate*} 

The node survival probability may be also interpreted as the cost needed to remove a node; thus, our findings offer an opportunity to understand which nodes have to be targeted/protected to deactivate a system or to keep it alive. Specifically, if node failure probability is constant across all nodes as in the Uniform model, then the best strategy always consists of attacking/protecting large degree nodes and such a conclusion echoes other findings in the literature~\cite{lu2016vital}. In contrast, if nodes display highly heterogeneous levels of resistance to failures (as in the BC model), then one has to carefully assess the trade-off between the costs and the corresponding connectivity loss deriving from node removal before taking a decision. 

Last but not least, significant differences emerged through the comparison (Sect.~\ref{sec:implications}) between our models with the state-of-art (i.e., the ``Benchmark'' analysis). Indeed, there is a deviation range that varies from 2\% up to 80\% regardless of the dataset used that highlight the existence of a significant gap between the common practice with a more realistic approach.

The paper is organized as follows: Section \ref{sec:background} provides some background materials, whereas in Section \ref{sec:network-reliability} our methodology is proposed, detailing the Uniform and the BC models. Section \ref{sec:related-literature} compares our approach with the related literature, whereas Section \ref{sec:experiments} illustrates the experimental findings. In Section \ref{sec:implications} the implications of this study are illustrated.
Finally, in Section \ref{sec:conclusions} the conclusions are drawn.

\section{Background}
\label{sec:background}

In this section, basic terminology from graph theory (Section \ref{sub:basic-terminology}) is introduced as well as the notion of node centrality in graphs (Section \ref{sub:node-centrality-networks}). Section \ref{sub:spectral-graph-role} relates on the graph spectral radius, and Section \ref{sub:component-graph-role} describes the role of the largest connected component in assessing graph robustness.

\subsection{Basic terminology on graphs}
\label{sub:basic-terminology}

A graph $G$ is a pair $ G = \langle N, E \rangle$ in which $N$ is the set of {\em nodes}, and $E = \{\langle i, j \rangle: i \in N \wedge j \in N\}$ is the set of {\em edges}.
Throughout the paper, the number of nodes (resp., edges) in $N$ (resp., $E$) are denoted by $n$ (resp., $2m$). We say that $G$ is {\em sparse}~\cite{Newman10} (resp., {\em dense}) if $m \in O(n)$ (resp., $m \in O(n^2)$), and is {\em undirected} if the edges are unordered pairs of nodes, or is {\em directed} otherwise. Our work focuses only on undirected graphs.

Given a node $i$, the neighbourhood $\mathcal{N}(i)$ of $i$ is the set of nodes connected to $i$.
The {\em degree} $d_i$ of $i$ is the number of edges incident onto $i$ (i.e., $d_i = \vert \mathcal{N}(i) \vert$). A {\em walk} of length $r>0$ is a sequence of alternating nodes and edges $i_1,e_1,i_2,e_2,\ldots,e_r,i_{r+1}$, such that for each $\ell = 1, \ldots, r$ the edge $e_{\ell}$ is $\langle i_{\ell}, i_{\ell+1} \rangle$. A {\em path} is a walk with no-repeated nodes.
A graph is {\em connected} if every pair of nodes are connected through a path.

Any graph is associated with a square matrix $\mathbf{A}$ -- called {\em adjacency matrix} --
such that $\mathbf{A}_{ij}$ is equal to $1$ if and only if there is an edge from node $i$ to node $j$; $0$, otherwise. 

The adjacency matrix of an undirected graph is {\em symmetric}, all its eigenvalues $\lambda_1 \geq \lambda_2 \geq \ldots \geq \lambda_n$ are {\em real}, and the corresponding eigenvectors $\mathbf{v}_1, \ldots, \mathbf{v}_n$ will form an {\em orthonormal basis} in $\mathbf{R}^n$~\cite{strang1993introduction}. 
The largest eigenvalue $\lambda_1$ of $\mathbf{A}$ is also called {\em spectral radius}.

\subsection{Node centrality in graphs}
\label{sub:node-centrality-networks}

The notion of node centrality has been introduced in late 1940s to quantify the importance of an actor in a social network~\cite{Newman10}. 
Roughly speaking, a {\em centrality metric} is a function $\phi: N \rightarrow \mathbb{R}^{+}$ that maps a node $i$ of a graph onto a real non-negative number $\phi(i)$, under the assumption that the larger $\phi(i)$, the more important $i$.

Some centrality metrics such as {\em degree}, {\em h-index}, and {\em coreness} depend on the ability of a node to influence its surrounding neighbours; hence, the most straightforward node centrality metric is the {\em degree}, which can be expressed as:

\begin{equation}
\label{eqn:degree}
\mathbf{d} = \mathbf{A} \times \mathbf{1}\
\end{equation}

Here $\mathbf{d}$ is a vector whose $i$-th entry denotes the degree of node $i$, $\mathbf{1}$ is the vector whose entries are all equal to $1$, and the symbol $\times$ denotes the usual matrix-by-matrix (or matrix-by-vector) product. 

The h-index, or {\em Hirsch index} (as the name of its creator), is a parameter that quantifies the academic impact of scientists~\cite{hirsch2005hindex}. For our purposes, the h-index is a local centrality metric defined as follows: a node $i$ in a graph $G$ has h-index $h$ if $i$ has at least $h$ neighbours, each of them with degree greater than or equal to $h$.

The {\em coreness}~\cite{lu2015hindex} of a node is grounded on the computation of the $\zeta$-{\em core} $G_{\zeta} = \langle N_{\zeta}, E_{\zeta} \rangle$ of a graph $G = \langle N, E \rangle$, being $\zeta$ a positive integer. Formally, the graph $G_\zeta$ is a subgraph of $G$, which satisfies the following properties: 
\begin{enumerate*}[label=(\alph*)]
\item each node in $N_{\zeta}$ has degree of at least $\zeta$, and\label{itm:1}
\item the graph $G_{\zeta}$ is {\em maximal} against property~\ref{itm:1}; i.e., if the aim is to add any node $j \in N - N_{\zeta}$ to $G_{\zeta}$ along with its edges, then the property~\ref{itm:1} would no longer hold true.
\end{enumerate*} 
Based on this definition, it may be asserted that a node $i \in N$ has coreness $\zeta$ if it belongs to $G_{\zeta}$, and it does not belong to $G_{\zeta +1}$.
Both the h-index and the coreness of a node are clearly related to the degree of that node, as reported in~\cite{lu2015hindex}.

Other centrality metrics -- e.g. the {\em Eigenvector} centrality~\cite{bonacich1987power} and the {\em Katz} centrality~\cite{katz1953new} -- rely on the full knowledge of the graph topology. 
Given a constant $\lambda \neq 0$, the Eigenvector centrality $\mathbf{e}$ is defined as the solution to the following equation:

\begin{equation}
\label{eqn:eigenvector-matrix}
\mathbf{A}\mathbf{e} = \lambda\mathbf{e}
\end{equation} 

If $G$ is connected and $\lambda = \lambda_1$, then the Perron-Frobenius Theorem~\cite{Newman10} states that there is a unique solution with all components positive to Equation \ref{eqn:eigenvector-matrix}, which corresponds to the largest eigenvector of $\mathbf{A}$. 

The Katz centrality, $\mathbf{k}$ is defined as follows:

\begin{equation}
\label{eqn:katz}
\mathbf{k} = \left(\mathbf{I} - \alpha\mathbf{A}\right)^{-1}\times \mathbf{1}
\end{equation}

Herein, $\mathbf{I}$ is the identity matrix and $\alpha$ is a fixed parameter that must be smaller than $1 / \lambda_1$.
If $\alpha \simeq 0$, then Katz centrality well approximates the degree. 
On the other hand, if $\alpha \simeq \frac{1}{\lambda_1}$, then Katz centrality is a good approximation of the Eigenvector centrality~\cite{benzi2014matrix,DeMeo19}. 
The semantics of the Katz centrality is as follows: given a node $i$, let consider all walks of arbitrary length starting from $i$ and ending in any other node $j$.
Node $i$ is assumed to be important if it is well connected to any other node through walks of arbitrary length. Yet, shorter walks have to be preferred to longer ones. To this purpose, walk length is weighted through a decreasing factor $\alpha$.

\subsection{The spectral radius of a graph and its role in governing dynamic processes}
\label{sub:spectral-graph-role}

The largest eigenvalue $\lambda_1$ of the adjacency matrix of a graph $G$ -- also known as the {\em spectral radius} -- 
can be used to analyze dynamical processes taking place over $G$, such as: the spread of a flu-like epidemics over a population or the spread of a malware in a computer network~\cite{chen2016node,wang2003epidemic,kleinberg2007computing}.

Early studies on virus propagation in human population pointed out the existence of a threshold $R_0$ (called {\em virus reproduction number}) such that if $\lambda_1 \geq R_0$ then a virus causes a global pandemics; whereas if $\lambda_1 < R_0$ the virus gets wiped out~\cite{hethcote2000mathematics,wang2003epidemic,ganesh2005effect}.

Due to its practical relevance, many authors were interested in assessing how $\lambda_1$ varies upon the removal of a target node~\cite{restrepo2006characterizing,tong2012gelling}. Specifically, Tong {\em et al.}~\cite{tong2012gelling} introduced the $k$-{\em node deletion problem}, which can be stated as follows: given an undirected graph $G = \langle N, E \rangle$, find the set of nodes $\mathcal{S}^{\star}(k) \subseteq N$ of cardinality $k$ which, if deleted from $G$, yield the biggest drop in $\lambda_1$.
The $k$-node deletion problem is NP-Hard~\cite{tong2012gelling}, thus efficient but accurate approximation algorithms are required to solve it. The state-of-the-art solution to the $k$-node deletion problem is the {\em NetShield} algorithm~\cite{chen2016node}, which achieves a worst-case time complexity of $O(nk^2 + m)$, being $n$ and $m$ the number of nodes and edges in $G$, respectively.

\subsection{The largest connected component of a graph}
\label{sub:component-graph-role}
Early studies investigated the decrease in network connectivity due to the selective removal (also known as {\em attacks}) of some of the network nodes or edges~\cite{barabasi1999emergence,holme2002attack,agreste2016network,albert2004structural}. 
An interesting class of attacks consists of repeatedly increasing the number of nodes/edges deleted from a graph $G$. This operation implies that $G$ breaks into disconnected subgraphs; thus, an important parameter to assess the ability of $G$ to preserve its functionality is given by the size $c$ of its largest connected component (LCC), i.e., the largest connected subgraph in $G$ after node/edge removal.

Studies in the field of OSNs indicate that $c$ is in the same order of magnitude of
the the entire network; thus, the LCC is also called {\em giant component}~\cite{holme2002attack,bollobas2001random}. 
Studies on the LCC size are also closely related to the topic of {\em percolation} and to the structure of random graphs~\cite{Newman10}. For instance, if an Erd\H os-R\' enyi random graph of $n$ nodes is considered, in which edges are placed uniformly at random between pair of nodes with probability $p_e$, then~\cite{bollobas2001random} proved that there exists a constant $\Psi$ such that if $p_e \geq p_e^{\star} = \frac{\left(1 + \Psi\right)}{n}$, and there exists a giant component in $G$ containing $O\left(n^{\frac{2}{3}}\right)$ nodes. On the contrary, if $p_e < p_e^{\star}$, then all the connected components of $G$ have the average size of $O(\log n)$.

We define the {\em transient phase} as the step in which $G$ moves from a highly-connected state to a new one in which the removal of a sufficiently high number of nodes leads to a significant decrease in the LCC size. The fragmentation process deriving from node removal is not gradual; it is characterized by a critical threshold $f_c$. If the fraction $f$ of removed nodes is less than $f_c$, then a giant component persists; but, once $f \geq f_c$, the giant component vanishes~\cite{Newman10,bollobas2001random}.

\section{A probabilistic node failure model}\label{sec:network-reliability}

This section presents our probabilistic node failure model, and the proposed protocol aimed to analyze both node failure impact on the spectral radius $\lambda_1$ and the largest connected component (LCC) size $c$ of a graph $G$. Let $\lambda_1$ (resp., $c$) be the spectral radius (resp., the LCC size of $G$).

Our methodology consists of the following components: 
\begin{enumerate*}[label=(\roman*)]
\item an undirected graph $G = \langle N, E \rangle$.
\item A {\em node-scoring} function $\phi: N \rightarrow \mathbb{R}^{+}$, which takes a node $i$ as input and returns its centrality $\phi(i)$ as output. Herein, the degree, h-index, coreness, Eigenvector, and Katz centrality are evaluated.
\item A {\em survival probability} function $\psi: N \rightarrow [0, 1]$, which takes a node $i$ as input and returns the probability $\psi(i)$ that $i$ will survive a failure.
\item A {\em target threshold} $\tau \in [0, 1]$, which specifies the fraction of nodes subject to failure.
\end{enumerate*}

Our methodology comprises the following steps:
\begin{enumerate}
 \item Each node $i \in N$ is associated with a score $\sigma_i = \left(1 - \psi(i)\right)\phi(i)$. Herein, high-score nodes are those with a high centrality (encoded in the factor $\phi(\cdot)$) and with a large probability of failing (expressed as
 $1 - \psi(\cdot)$).
 %, because only failed nodes actually contribute to the loss of connectivity.
 \item The top $\lceil \tau \vert N_0 \vert \rceil$ nodes with largest scores are picked and deleted from $G$ along with their edges.
\end{enumerate}

The procedure above yields a new graph $\tilde{G}(\tau, \phi)$ with spectral radius $\tilde{\lambda}_1(\tau, \phi)$ and the LCC size $\tilde{c}(\tau, \phi)$.

Two variants of {\em probabilistic node failure model}, namely {\em Uniform} and {\em Best Connected}(BC) are considered.
In the {\em Uniform} model, it supposes that $\psi(i) = p$ for every node $i$, being $p$ a fixed value in $[0, 1]$. The BC model is grounded on the principle that nodes display an unequal level of tolerance to failures: intuitively, large degree nodes have to occupy a prominent position in $G$ because their removal may quickly lead to network fragmentation~\cite{albert2000error}; thus, they should display a better resistance to failures. 
A possible model of $\psi(\cdot)$ -- which incorporates the observations above -- is $\psi(i) = \frac{d_i}{2m}$.
Other models to describe node resistance to failures are also allowed, but we leave their discussion as future work.

Two parameters are introduced, namely the {\em effectiveness} and the {\em coverage} to quantify the loss in connectivity that $G$ suffers upon node removal:

\begin{definition}
Let $G$ be an undirected and connected graph and let $\tau \in [0, 1]$.
Let $\tilde{G}(\tau, \phi)$ be the graph obtained from $G$ by applying the probabilistic node failure model above with $\phi$ as centrality metric and $\tau$ as target threshold.

The {\em effectiveness} $\rho(\tau, \phi)$ of $\phi$ in the Uniform (resp., BC) model is defined as:
\begin{equation}
\label{eqn:rho}
\rho(\tau, \phi) = \frac{\tilde{\lambda}_1(\tau, \phi)}{\lambda_1}
\end{equation}

where $\tilde{\lambda}_1(\tau, \phi)$ (resp., $\lambda_1$) is the spectral radius of $\tilde{G}(\tau, \phi)$ (resp., $G$).

The {\em coverage} $\gamma(\tau, \phi)$ of $\phi$ in the Uniform (resp., BC) model is defined as:
\begin{equation}
\label{eqn:gamma}
\gamma(\tau, \phi) = \frac{\tilde{c}(\tau, \phi)}{c}
\end{equation}

where $\tilde{c}(\tau, \phi)$ (resp., $c$) is the LCC size of $\tilde{G}(\tau, \phi)$ (resp., $G$).
\end{definition}

Because of $\lambda_1(\tau, \phi) \leq \lambda_1$ (see~\cite{stewart1990matrix}), the effectiveness $\rho(\tau, \phi)$ always ranges in $[0, 1]$ and the closer $\rho(\tau, \phi)$ to zero, the more effective $\phi$.

Analogously, nodes removal from $G$ leads to a shrinkage in the LCC size observed in $\tilde{G}(\tau, \phi)$, which implies that $\gamma(\tau, \phi) \in [0, 1]$. The closer $\gamma(\tau, \phi)$ to zero, the higher the shrinkage of the largest connected component.

In Section \ref{sec:experiments} the variation of effectiveness and coverage is analyzed, through a set of experiments on real-life graphs.

\section{Related Literature}
\label{sec:related-literature}

This section reviews past research works related to this paper.
Firstly, the ability of centrality metrics to identify nodes in a graph that give rise and favour diffusion processes are described (Section \ref{sub:detecting-top-spreaders}). 
Then, in Section \ref{sub:topology-manipulation}, approaches that manipulate graph topology are reviewed, and investigated how these modifications alter centrality metrics, as well as other graph parameters.

\subsection{Identifying nodes capable of activating diffusion processes}
\label{sub:detecting-top-spreaders}

The problem of detecting nodes that originate diffusion processes in networks has been extensively studied in the past, and is well aligned with the problem of calculating centrality scores in graphs~\cite{lu2016vital,LoKa19}.
A relevant application is the study of the misinformation spreading in OSNs~\cite{comin2011identifying,jiang2017identifying}. 

For instance, Comin and da Fontoura Costa~\cite{comin2011identifying} applied standard centrality metrics like degree centrality to identify the sources of misinformation. 
Shah and Zaman~\cite{shah2011rumors} introduced an ad-hoc centrality parameter, called {\em \red{rumour} centrality}, to rank nodes on the basis of their spreading ability. They focused on tree-like networks and hypothesized that a node can receive information from only one of its neighbours. Dong {\em et al.}~\cite{dong2013rooting} extended the approach of Shah and Zaman~\cite{shah2011rumors} by detecting nodes with the largest \red{rumour} centrality within a set of suspected nodes. 

Contrary to the aforementioned approaches, we consider graphs of arbitrary topology and we are on a quest to assess how the spectral radius $\lambda_1$ and the LCC size $c$ of a graph vary upon the random failure of some nodes.

Prakash {\em et al.}~\cite{prakash2014efficiently} applied spectral methods to evaluate the spreading power of nodes. However, due to its high computational costs, their method is applicable only to small-size graphs. Nguyen {\em et al.}~\cite{nguyen2012sources} employed the Monte Carlo techniques to discover the set of nodes that are the best candidates for spreading misinformation.
Budak {\em et al.}~\cite{budak2011limiting} considered competing campaigns over a social network and aimed at identifying a subset of individuals that need to be convinced to promote a \lq\lq good\rq\rq campaign to minimize the number of people who adhere to a \lq\lq bad\rq\rq one. 
They proved that degree centrality is a good heuristic to find out nodes involved in good campaigns, provided that the delay elapsing between the start of misinformation spread and its first detection is fairly small.

These methods assume that some nodes in networks should be better protected to neutralize misinformation spread. Such a belief agrees with a core assumption of our approach: in fact, in the Best Connected (BC) model, we suppose that large degree nodes occupy a crucial role in the system functioning. Thus, they must display a larger survival probability to failures.

A relevant difference between approaches from the literature and ours is that the neutralization strategy requires to solve an optimization problem, whereas we are in charge of evaluating the deformation of $\lambda_1$ and $c$. In addition, as a guiding criterion to select nodes to remove, we adopt centrality metrics, which are easy to calculate and have a clear interpretation.

\subsection{Variation in graph connectivity after nodes removal}
\label{sub:topology-manipulation}

Node removal procedures have been applied to investigate the resilience of large systems~\cite{albert2000error,borgatti2006robustness,restrepo2006characterizing}. 
Albert {\em et al.}~\cite{albert2000error} studied how diameter and the size of the giant component of \red{Erd\H os-R\'enyi} and scale-free graphs varied when nodes were removed at random, or if large degree nodes were deleted.
Borgatti {\em et al.}~\cite{borgatti2006robustness} examined the accuracy in estimating centrality scores if graph data are incomplete. In the Web search domain, Ng {\em et al.}~\cite{ng2001link} examined small changes of the Web graph and their impact on the PageRank and HITS scores.

Restrepo {\em et al.}~\cite{restrepo2006characterizing} defined the {\em dynamical importance} of a node $i$ as the amount $-\Delta\lambda_{i}$ by which $\lambda_1$ decreases upon removal of the all edges incident onto $i$, normalized by $\lambda_1$. 
Al-Dabbagh~\cite{Al-Dabbagh19} studied the topology design of a wireless control system in which nodes and wireless links are unreliable (due, for instance, to battery drainage). The approach of~\cite{Al-Dabbagh19} (shared by us) assumes that network elements may fail in an unpredictable way; yet the focus is to determine whether it is possible to design a controller for a wireless network, given that the largest number of unreliable nodes in the network and the probability that a link fails are specified. 

Unlike approaches described in literature, we manage a scenario in which nodes may survive a failure; thus, we considered a probabilistic framework to describe node failure. 
In addition, many of the approaches discussed in this section fix the number $k$ of nodes/edges to be removed, and attempt at finding the best strategy to delete at most $k$ nodes. In contrast, we aim at experimentally studying the variation in $\lambda_1$ and $c$ when the fraction of nodes subject to failure varies.

\section{Experimental Analysis}
\label{sec:experiments}

This section shows the experiments carried out to evaluate the effectiveness and coverage of degree, h-index, coreness, Eigenvector and Katz centrality in both the Uniform and the Best Connected (BC) models.

\subsection{Datasets}
\label{sub:datasets}

Four real-life graphs to perform the experimental analysis have been employed, all taken from the Konect repository\footnote{\url{konect.uni-koblenz.de/networks/}}, namely: \textsc{US\_POWER\_GRID} (describing the power grid in US Western states), \textsc{AstroPh} (describing co-authorship between scientists in the astrophysics domain), \textsc{BrightKite} (a location-based social networking Web site) and \textsc{Flickr} (a graph whose nodes represent Flickr photos and an edge indicates that two photos share some tags).
The datasets features are summarized in Table \ref{tbl:dataset}.

\begin{table}
\centering
    \begin{tabular}{clrrl}
        \hline \hline 
        & Dataset & \# Nodes & \# Edges\\ %& $\lambda_1$ \\
        \hline 
        1 & \textsc{US\_Power\_Grid} & $4\,941$ & $6\,594$\\% & $7.48$ \\
        2 & \textsc{AstroPh} & $18\,771$ & $198\,050$\\% & $94.43$ \\
        3 & \textsc{BrightKite} & $58\,228$ & $214\,078$\\% & $101.49$ \\
        4 & \textsc{Flickr} & $105\,938$ & $2\,316\,948$\\% & $615.56$ \\
    \hline
    \end{tabular}
\caption{Datasets adopted in the experimental trials. For each dataset the number of nodes, and the number of edges are reported.} 
~\label{tbl:dataset}
\end{table}

\begin{figure*}[htb]
  \centering  
  \begin{subfigure}{0.245\textwidth}
    \includegraphics [width=\textwidth] {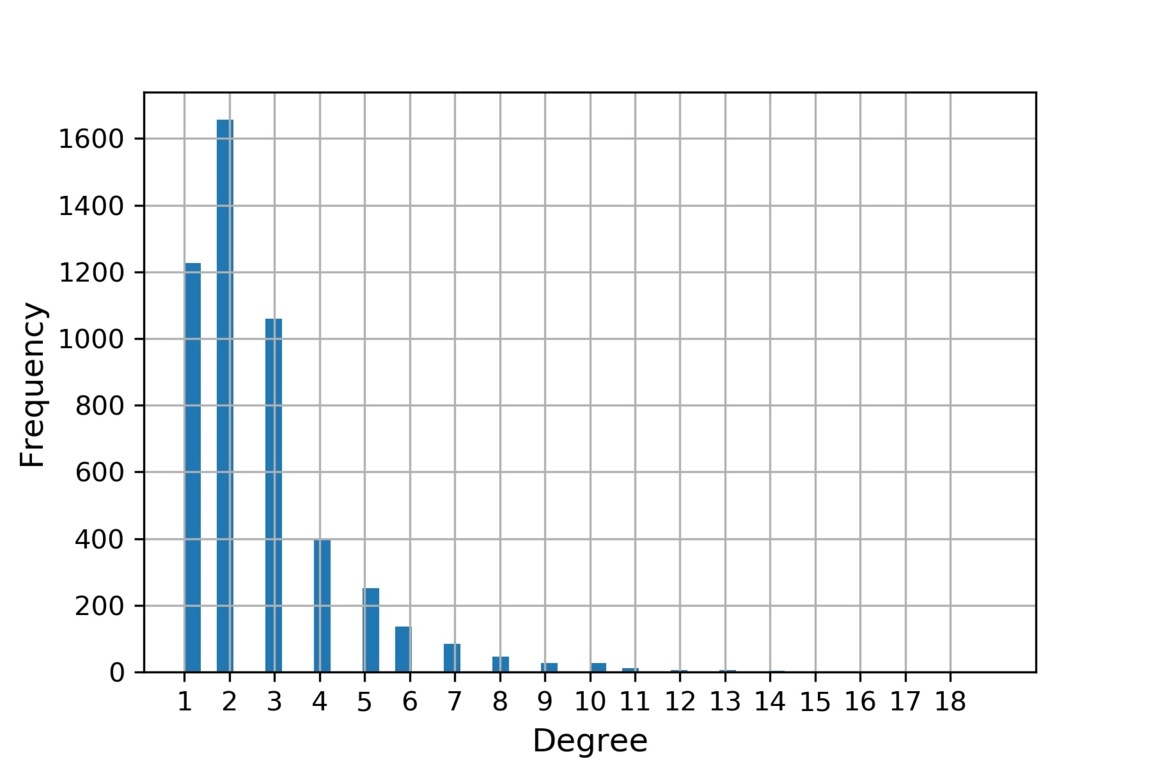}
    \vspace{-2em}
    \caption{\textsc{\textsc{US\_Power\_Grid}}}
  \end{subfigure}
  \begin{subfigure}{0.245\textwidth}
    \includegraphics [width=\textwidth] {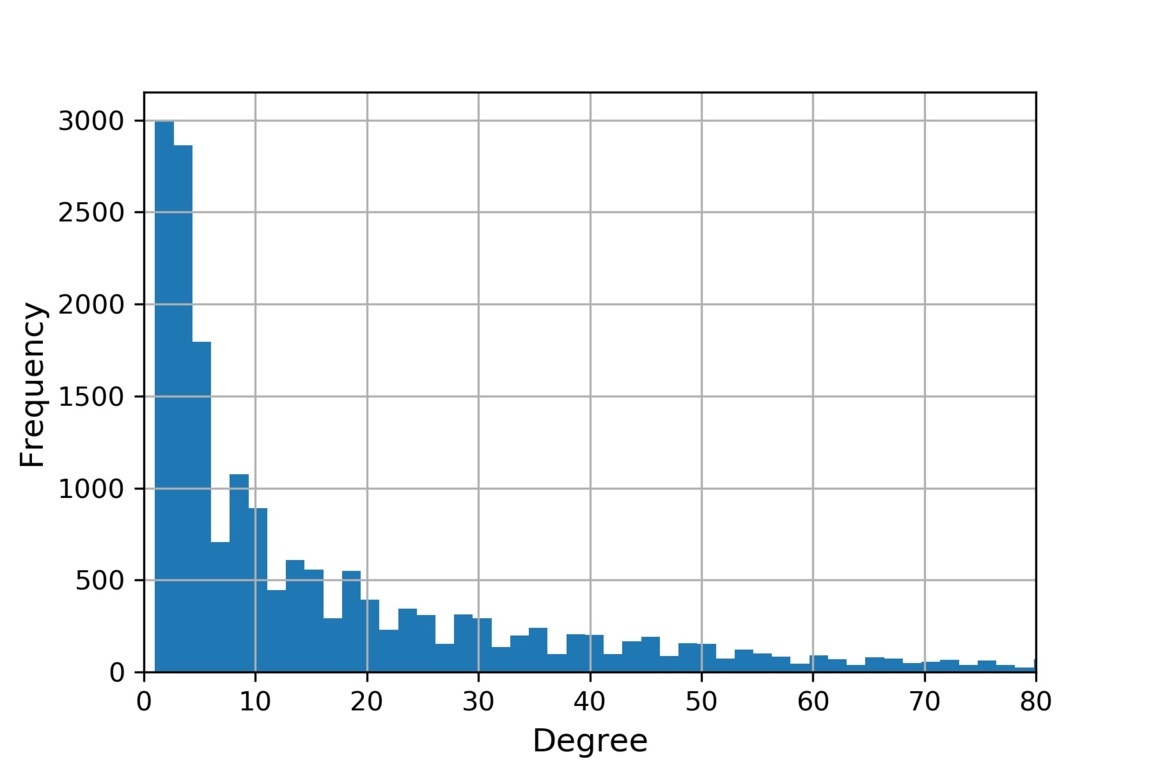}
    \vspace{-2em}
    \caption{\textsc{AstroPh}}
  \end{subfigure}
  \begin{subfigure}{0.245\textwidth}
    \includegraphics [width=\textwidth] {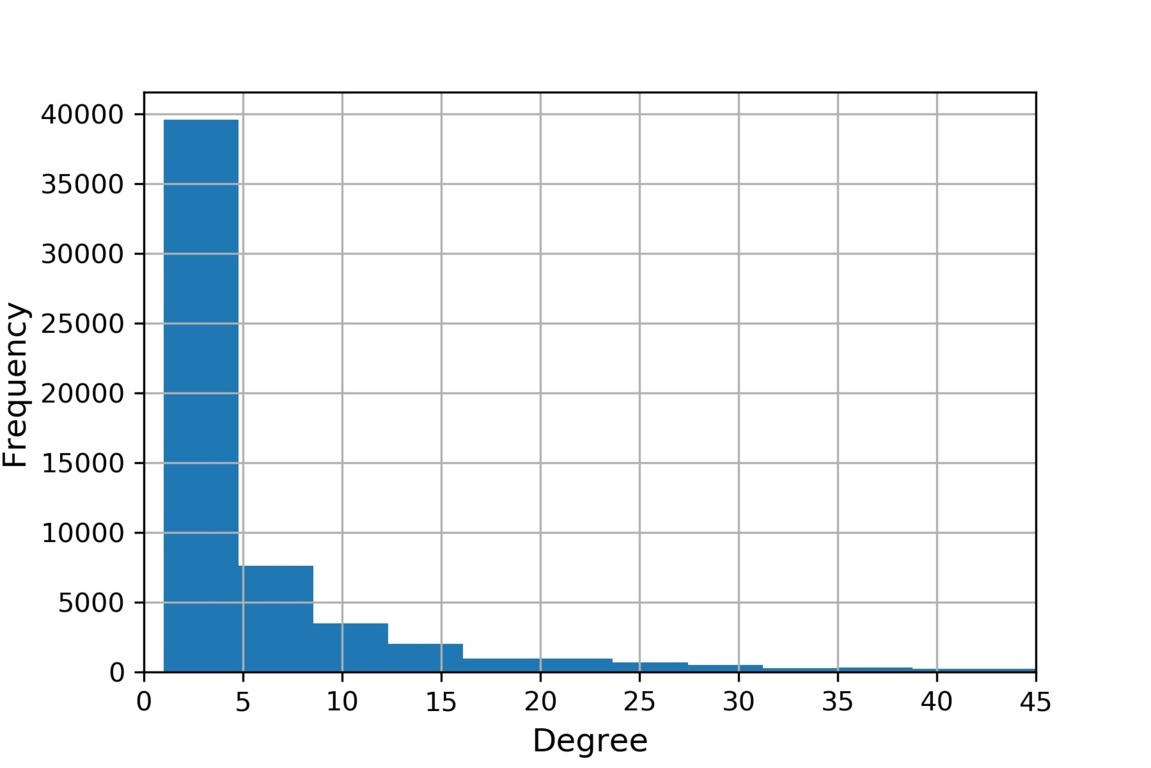}
    \vspace{-2em}
    \caption{\textsc{BrightKite}}
  \end{subfigure}
  \begin{subfigure}{0.245\textwidth}
    \includegraphics [width=\textwidth] {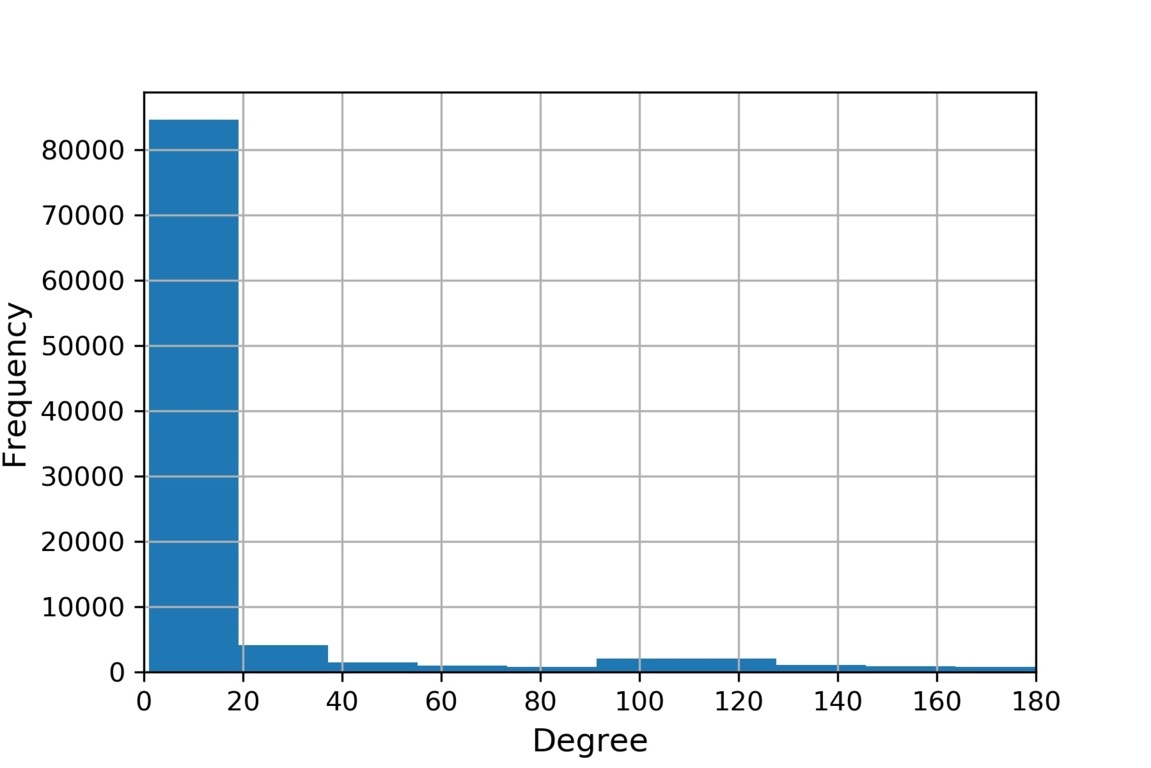}
    \vspace{-2em}
    \caption{\textsc{Flickr}}
  \end{subfigure}
    \vspace{-0.8em}
     \caption{Degree Distribution in the \textsc{US\_Power\_Grid}, \textsc{AstroPh}, \textsc{BrightKite} and \textsc{Flickr} datasets.}
\label{fig:degree-distribution}
\end{figure*}

In Figure \ref{fig:degree-distribution} the degree distribution of all graphs involved in this study is reported.

The degree distribution in \textsc{US\_Power\_Grid} significantly deviates from that observed in other graphs: herein, most of the nodes have a degree from one to three, but a non-negligible fraction of nodes with degree from four to seven has been highlighted. 
Node degree distribution in other graphs is highly skewed: for instance, in \textsc{BrightKite}, roughly 70\% of nodes has degree less than five and more than $80\%$ of nodes in \textsc{Flickr} display a degree less than twenty, although around one thousand nodes in the \textsc{Flickr} graph have a degree from 90 to 130.

\subsection{Experimental setup}
\label{sub:experimental-setup}

In the experiments, the fraction $\tau$ of nodes to remove from $0$ to $0.18$ has been varied. 
As for the Uniform model, three different values of the survival probability $p$, namely, $0.1$, $0.3$, and $0.5$ have been considered.

\begin{figure*}[htb]
  \centering
  \begin{subfigure}{0.245\textwidth}
    \includegraphics [width=\textwidth] {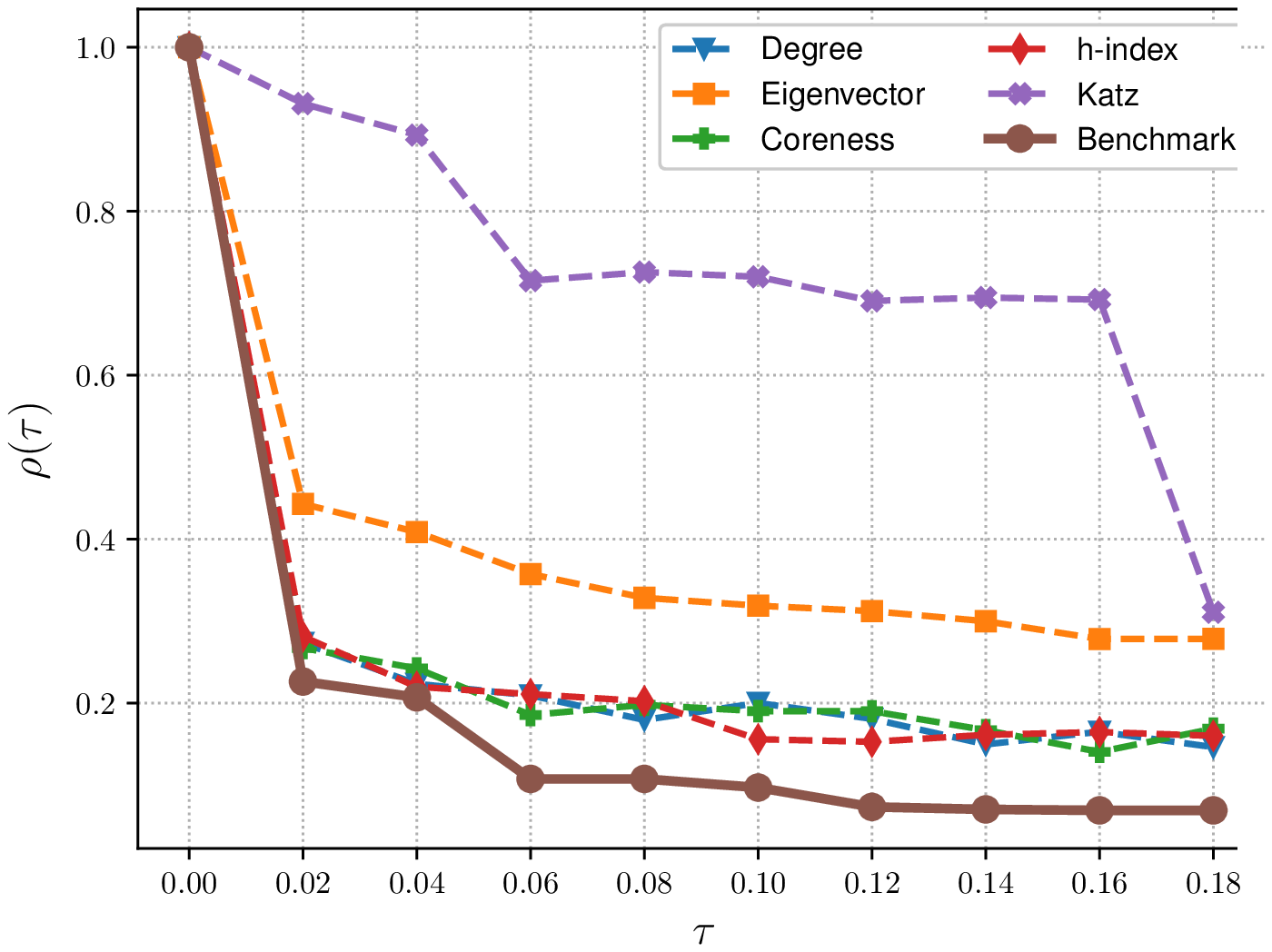}
    \vspace{-2em}
    \caption{Uniform $p=0.1$}
  \end{subfigure}
  \begin{subfigure}{0.245\textwidth}
    \includegraphics [width=\textwidth] {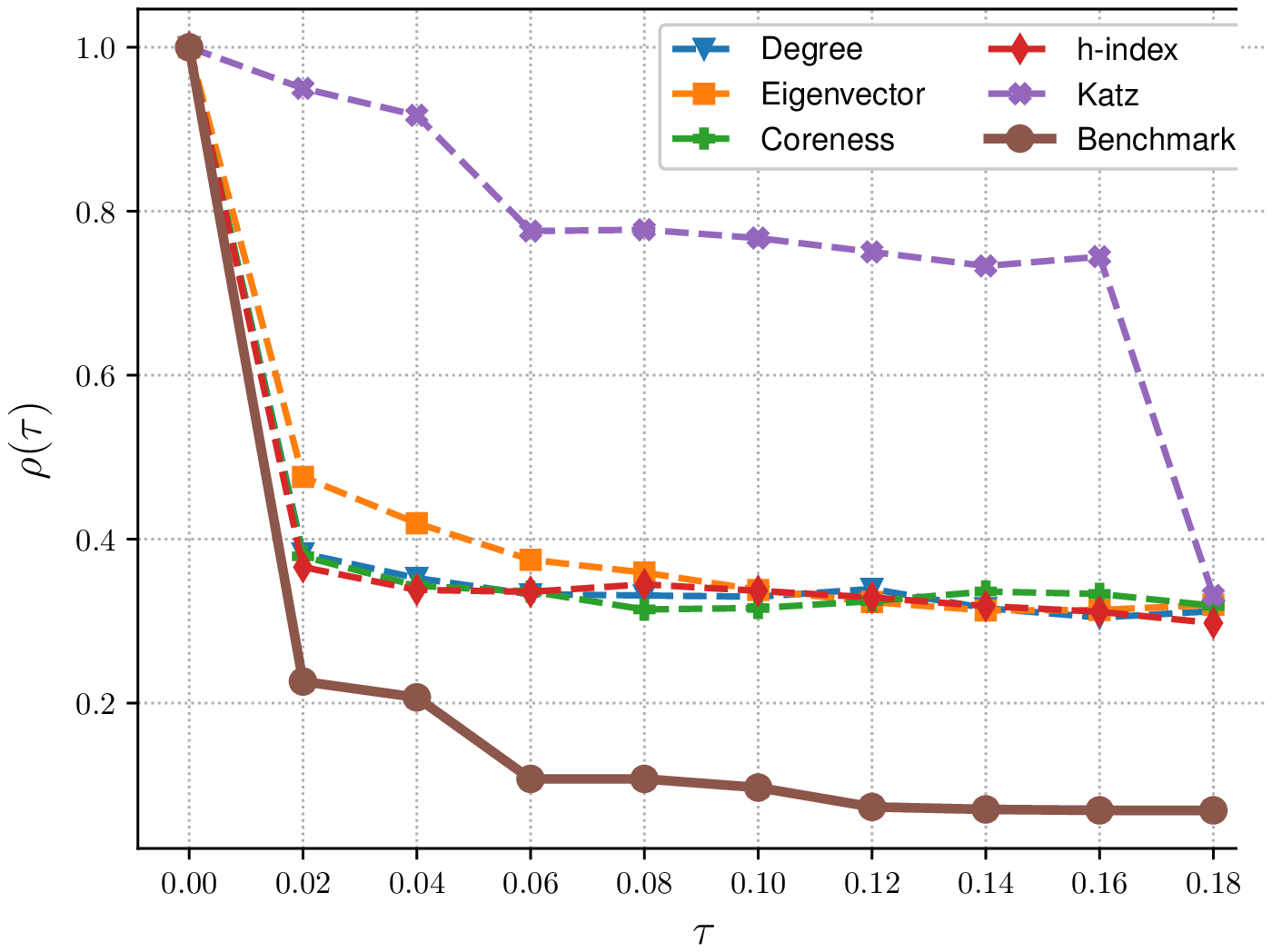}
    \vspace{-2em}
    \caption{Uniform $p=0.3$}
  \end{subfigure}
  \begin{subfigure}{0.245\textwidth}
    \includegraphics [width=\textwidth] {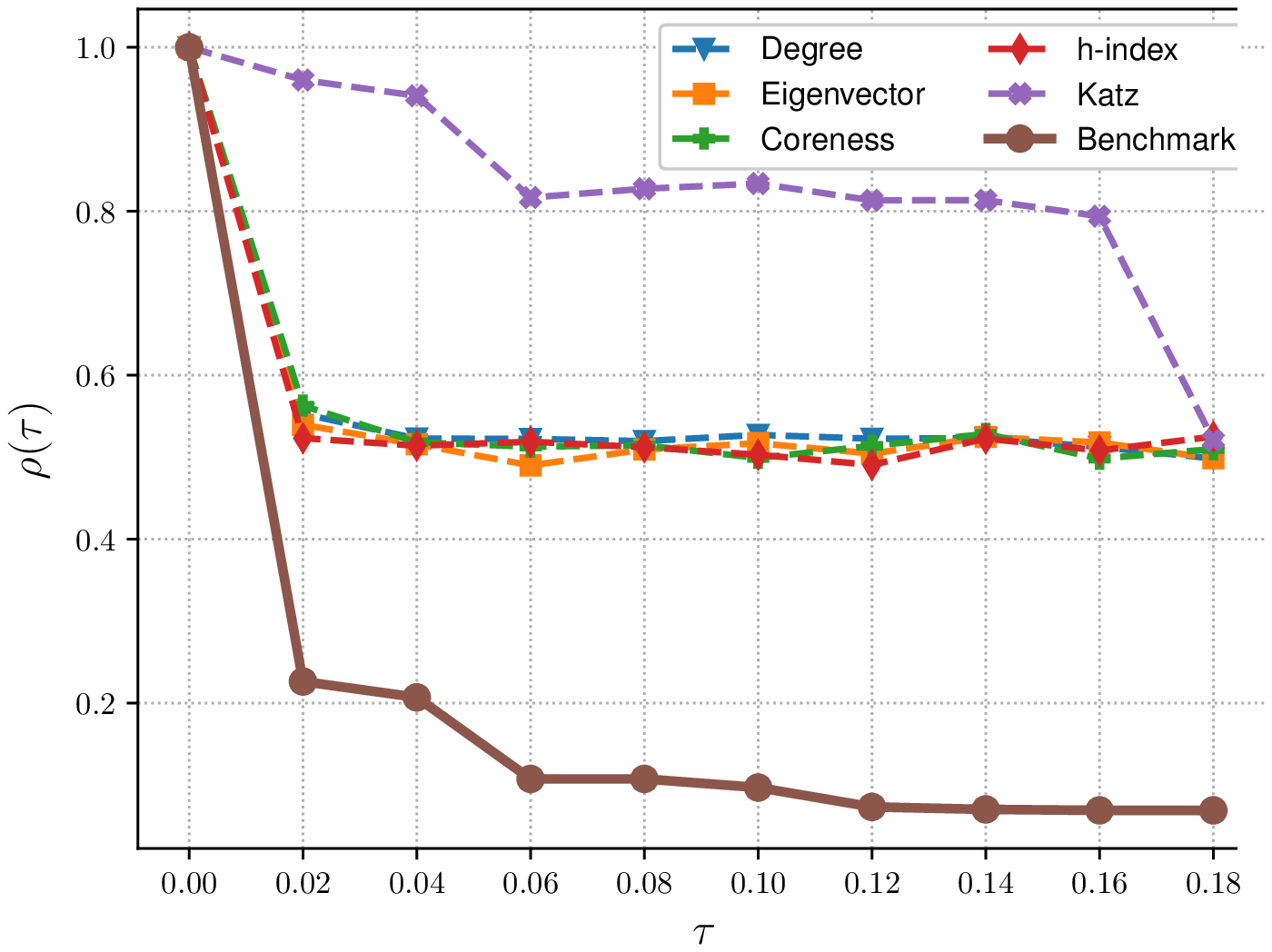}
    \vspace{-2em}
    \caption{Uniform $p=0.5$}
  \end{subfigure}
  \begin{subfigure}{0.245\textwidth}
    \includegraphics [width=\textwidth] {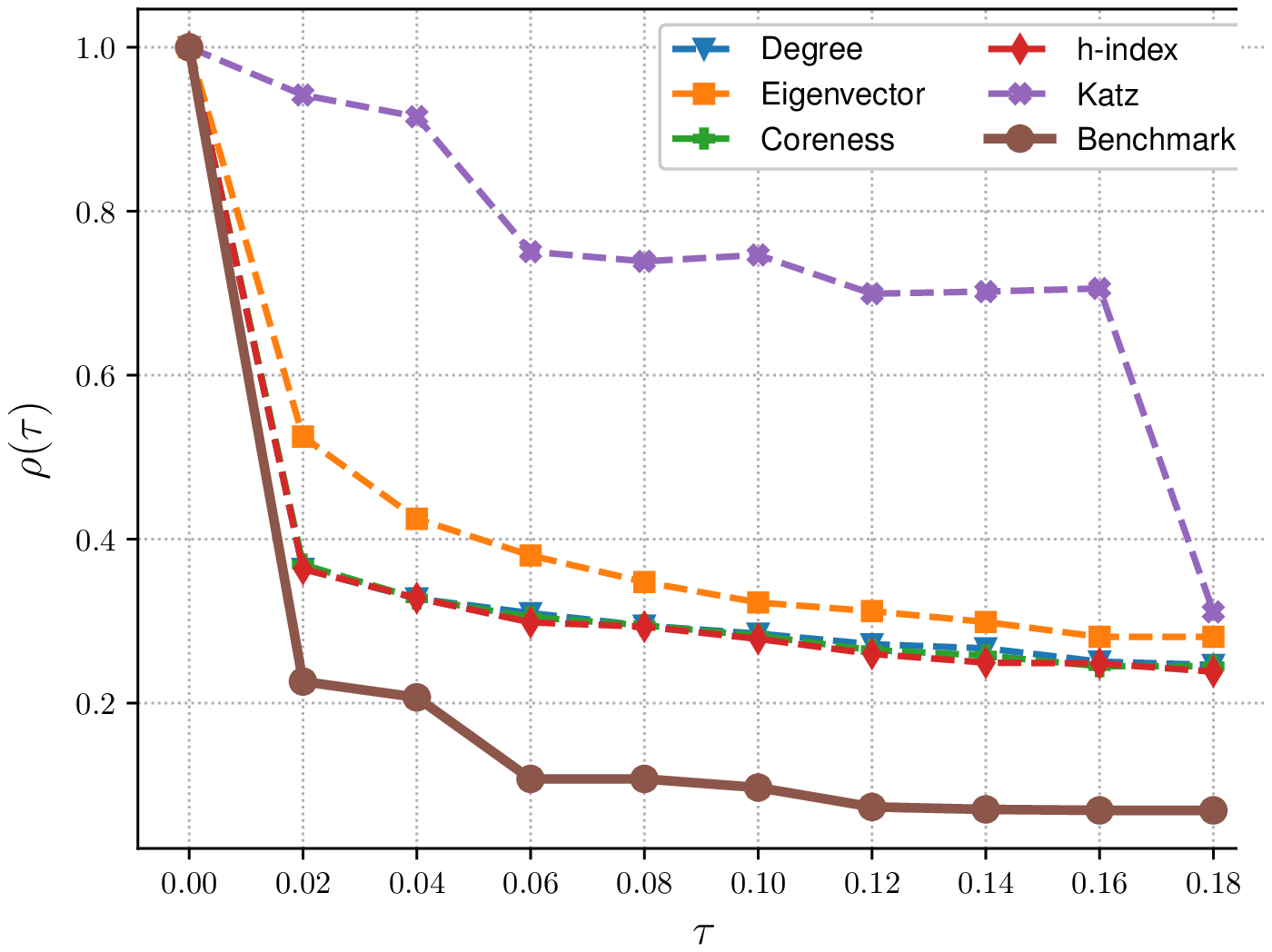}
    \vspace{-2em}
    \caption{BC}
  \end{subfigure}
    \vspace{-0.8em}
      \caption{Effectiveness tests on \textsc{Brightkite} dataset: (a-c) Uniform model with $p \in \{0.1,0.3,0.5\}$, (d) BC.}
  \label{fig:Brightkite}
\end{figure*}

\begin{figure*}[htb]
  \centering  
  \begin{subfigure}{0.245\textwidth}
    \includegraphics [width=\textwidth] {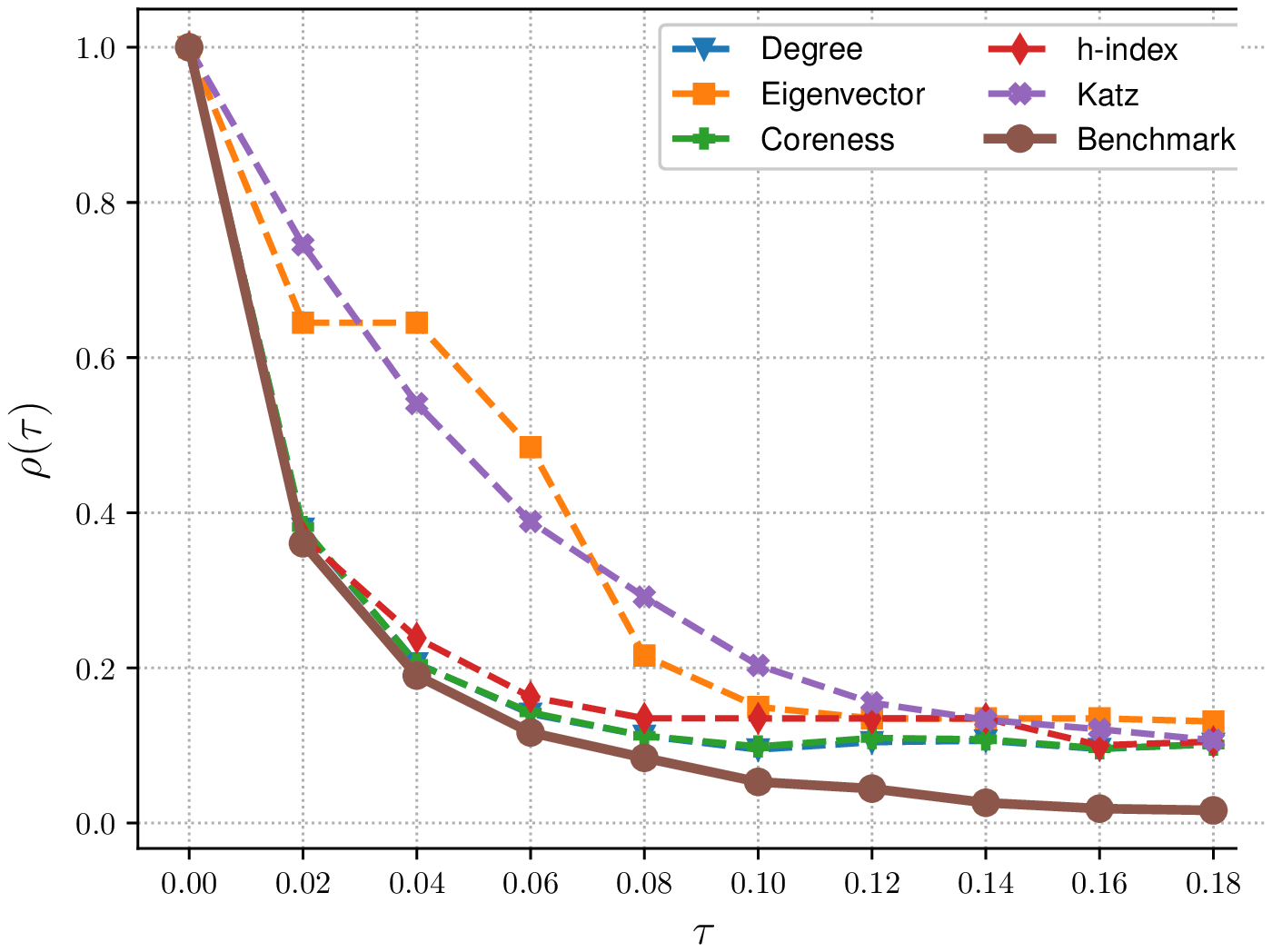}
    \vspace{-2em}
    \caption{Uniform $p=0.1$}
  \end{subfigure}
  \begin{subfigure}{0.245\textwidth}
    \includegraphics [width=\textwidth] {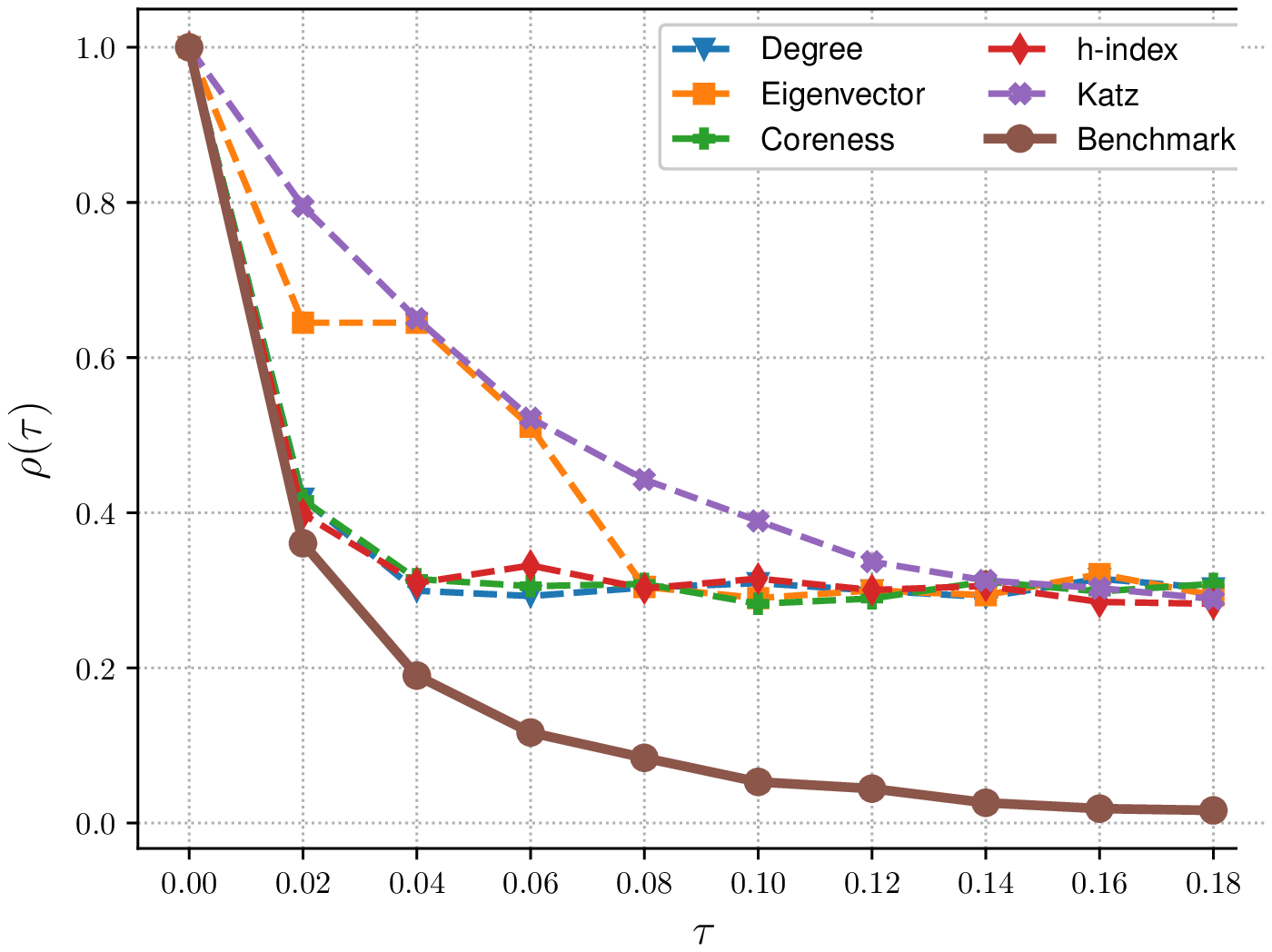}
    \vspace{-2em}
    \caption{Uniform $p=0.3$}
  \end{subfigure}
  \begin{subfigure}{0.245\textwidth}
    \includegraphics [width=\textwidth] {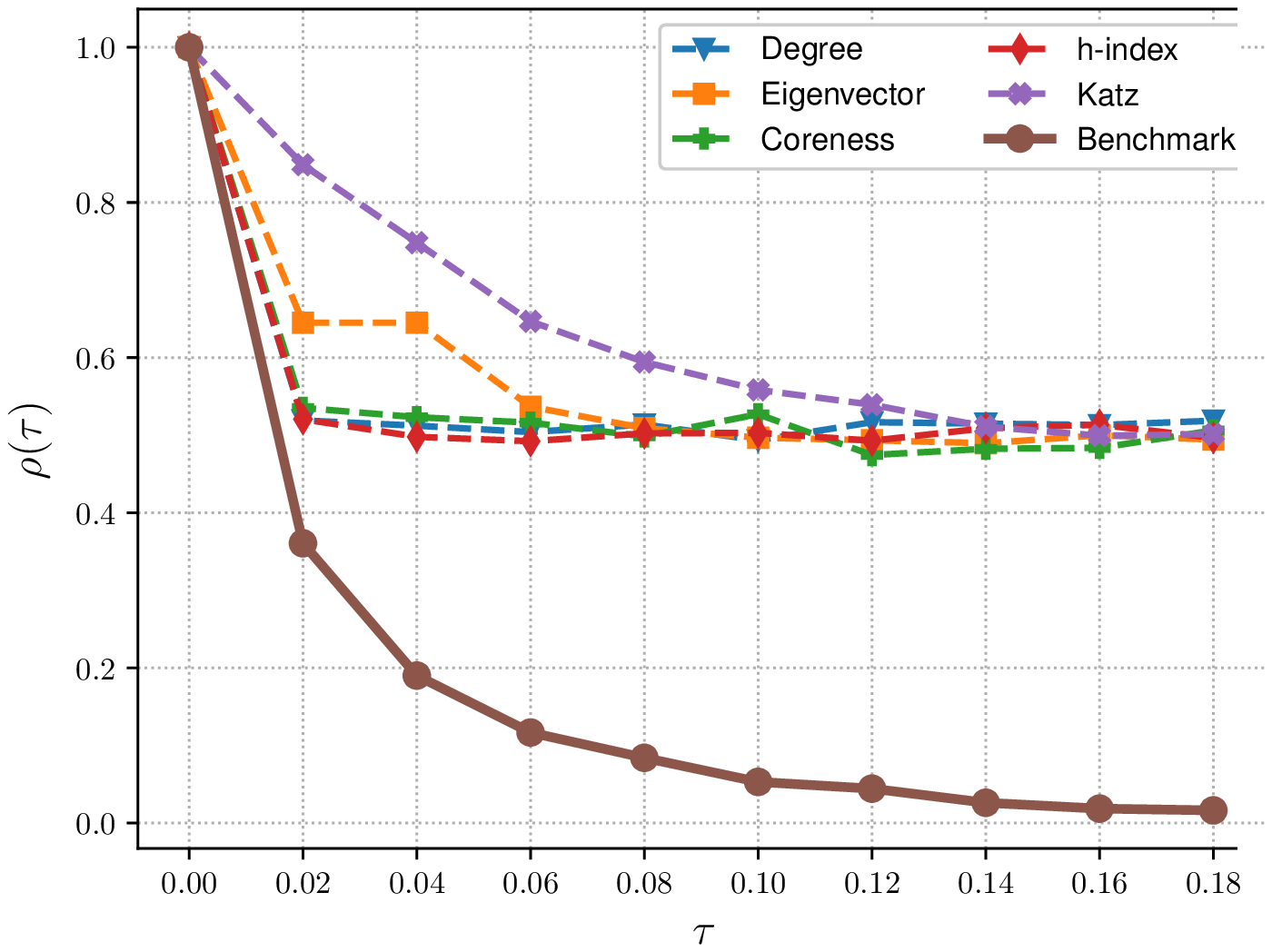}
    \vspace{-2em}
    \caption{Uniform $p=0.5$}
  \end{subfigure}
  \begin{subfigure}{0.245\textwidth}
    \includegraphics [width=\textwidth] {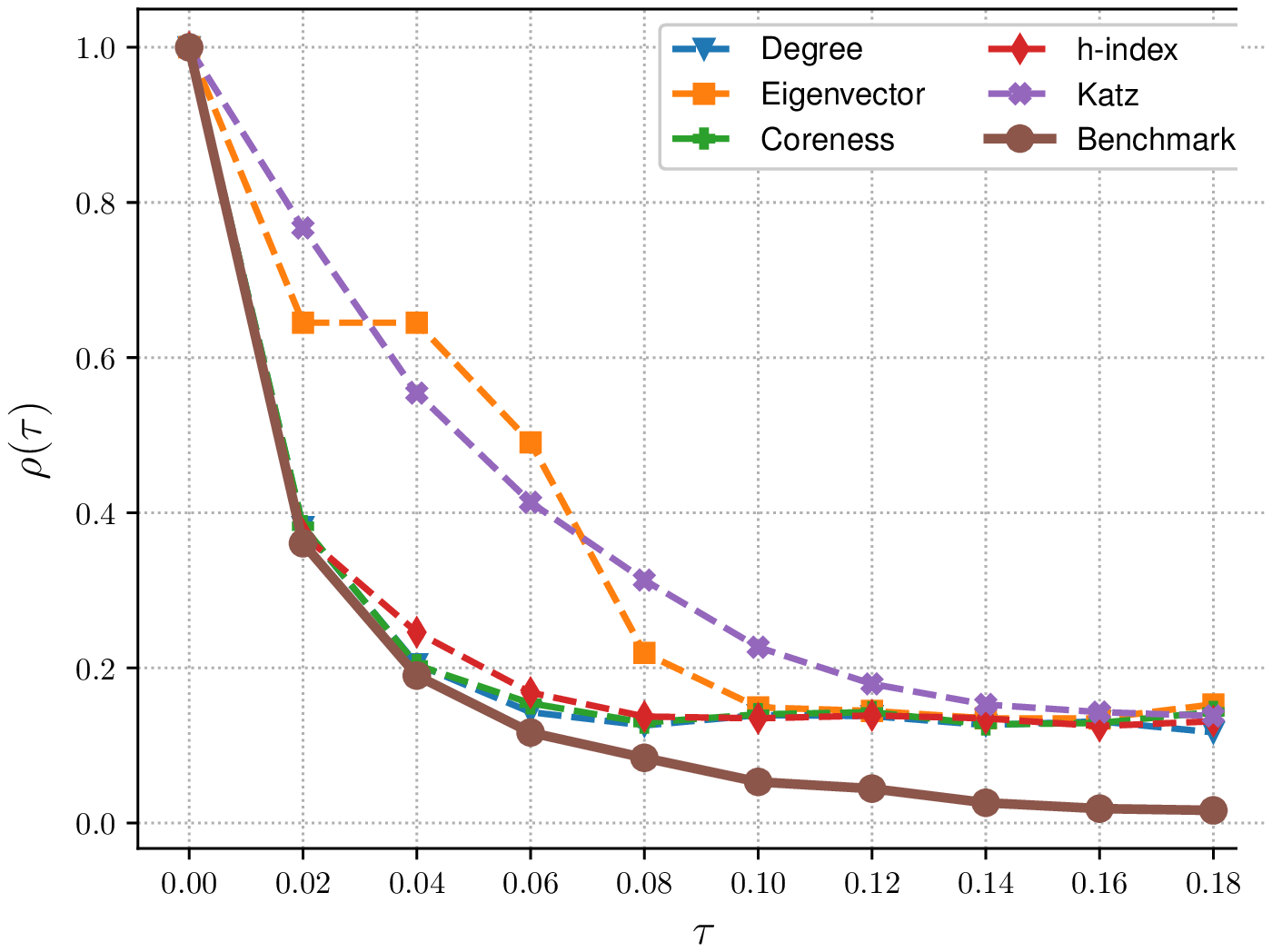}
    \vspace{-2em}
    \caption{BC}
  \end{subfigure}
    \vspace{-0.8em}
     \caption{Effectiveness tests on \textsc{Flickr} dataset: (a-c) Uniform model with $p \in \{0.1,0.3,0.5\}$, (d) BC.}
\label{fig:Flickr}
\end{figure*}

Note that in all the effectiveness and coverage plots (Figures \ref{fig:US-POWER_GRID}-\ref{fig:Flickr-cc}), also the Benchmark analysis has been reported. This curve represents the normal behaviour of nodes removal process if the probabilistic failure is not considered (i.e., as if $p=0$ in the Uniform model) by using the degree centrality metric that will be later confirmed to be the most effective one to analyse both $\lambda_1$ and $c$ drops. Thus, the Benchmark allows to make a comparison between the state-of-art approach and our more realistic ones.

Due to space limitations, only the results for the Katz centrality with $\alpha = 0.1$ have been reported.

The experiments' results have been averaged 20 times to avoid statistical fluctuations.

\subsection{Effectiveness of Centrality Measures}
\label{sub:effectiveness}

\begin{figure*}[htb]
  \centering  
  \begin{subfigure}{0.245\textwidth}
    \includegraphics [width=\textwidth] {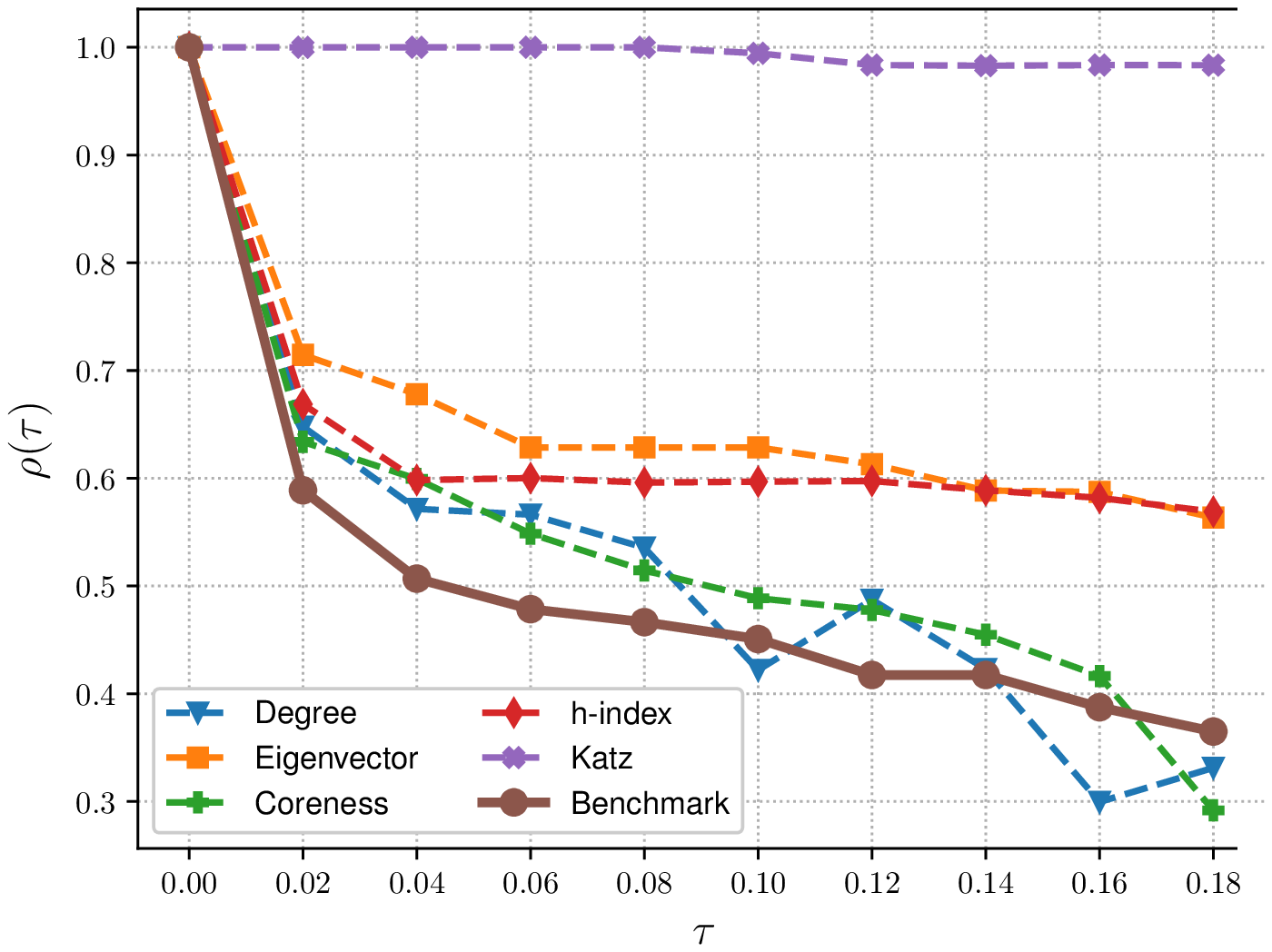}
    \vspace{-2em}
    \caption{Uniform $p=0.1$}
  \end{subfigure}
  \begin{subfigure}{0.245\textwidth}
    \includegraphics [width=\textwidth] {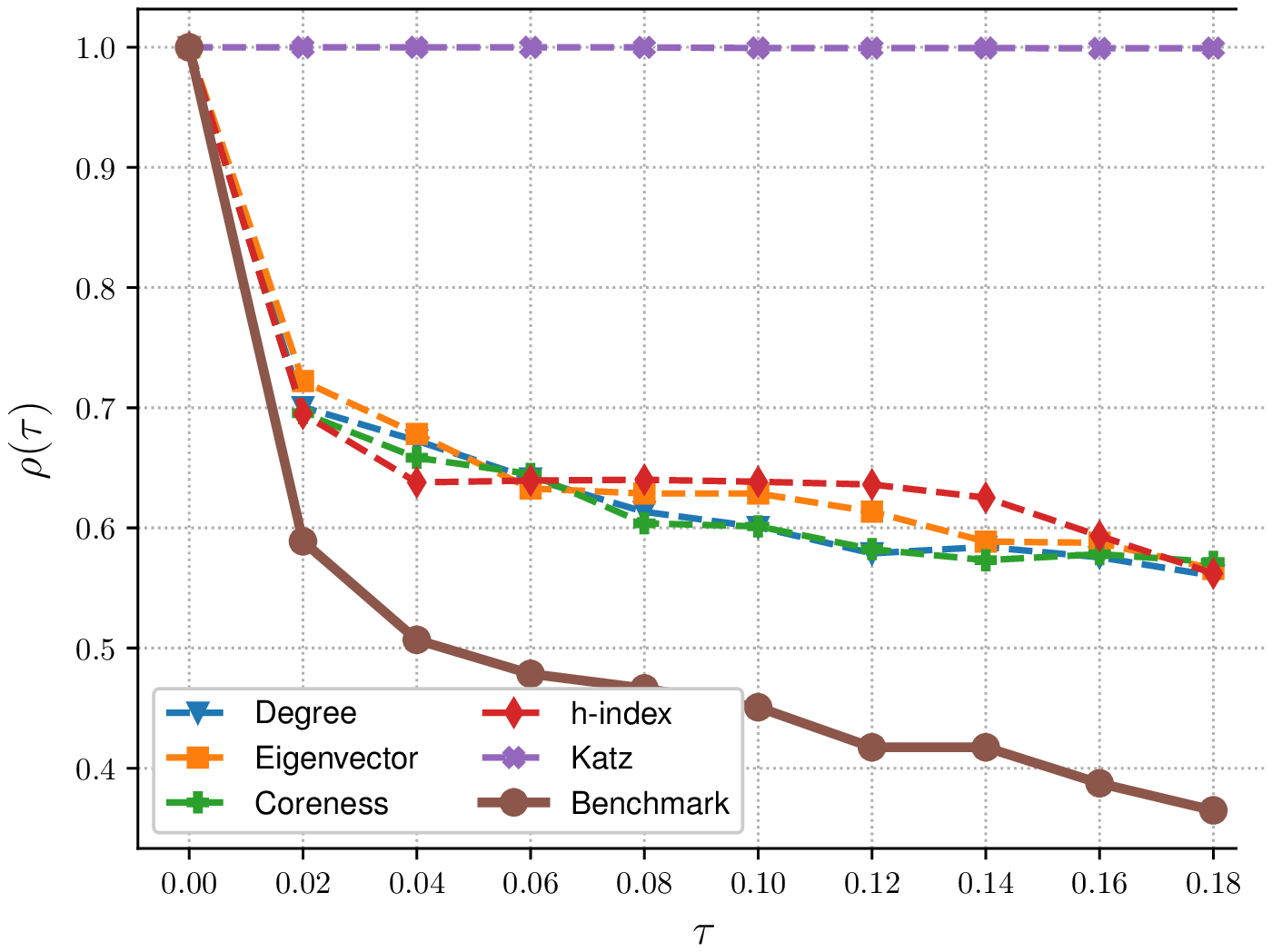}
    \vspace{-2em}
    \caption{Uniform $p=0.3$}
  \end{subfigure}
  \begin{subfigure}{0.245\textwidth}
    \includegraphics [width=\textwidth] {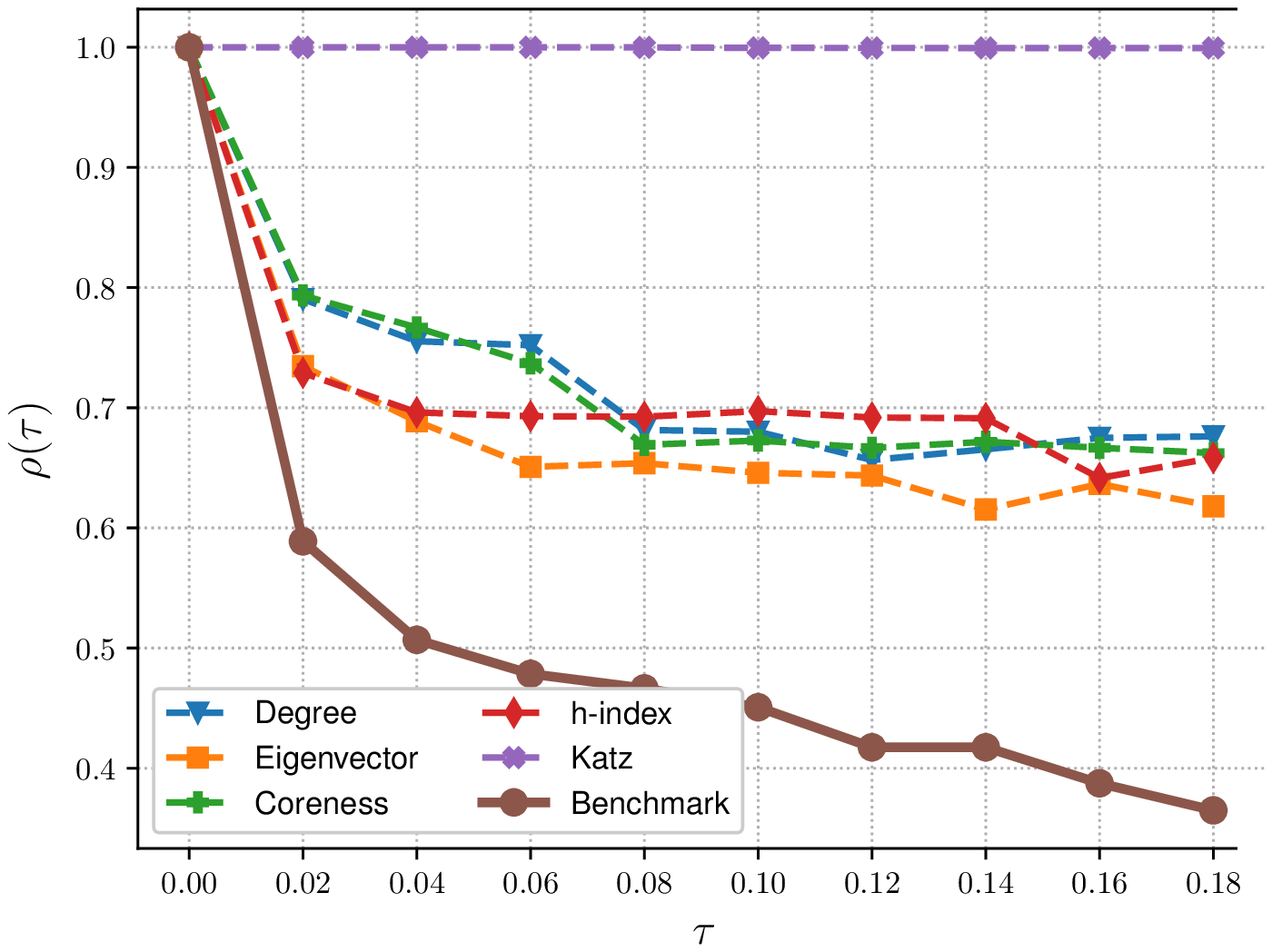}
    \vspace{-2em}
    \caption{Uniform $p=0.5$}
  \end{subfigure}
  \begin{subfigure}{0.245\textwidth}
    \includegraphics [width=\textwidth] {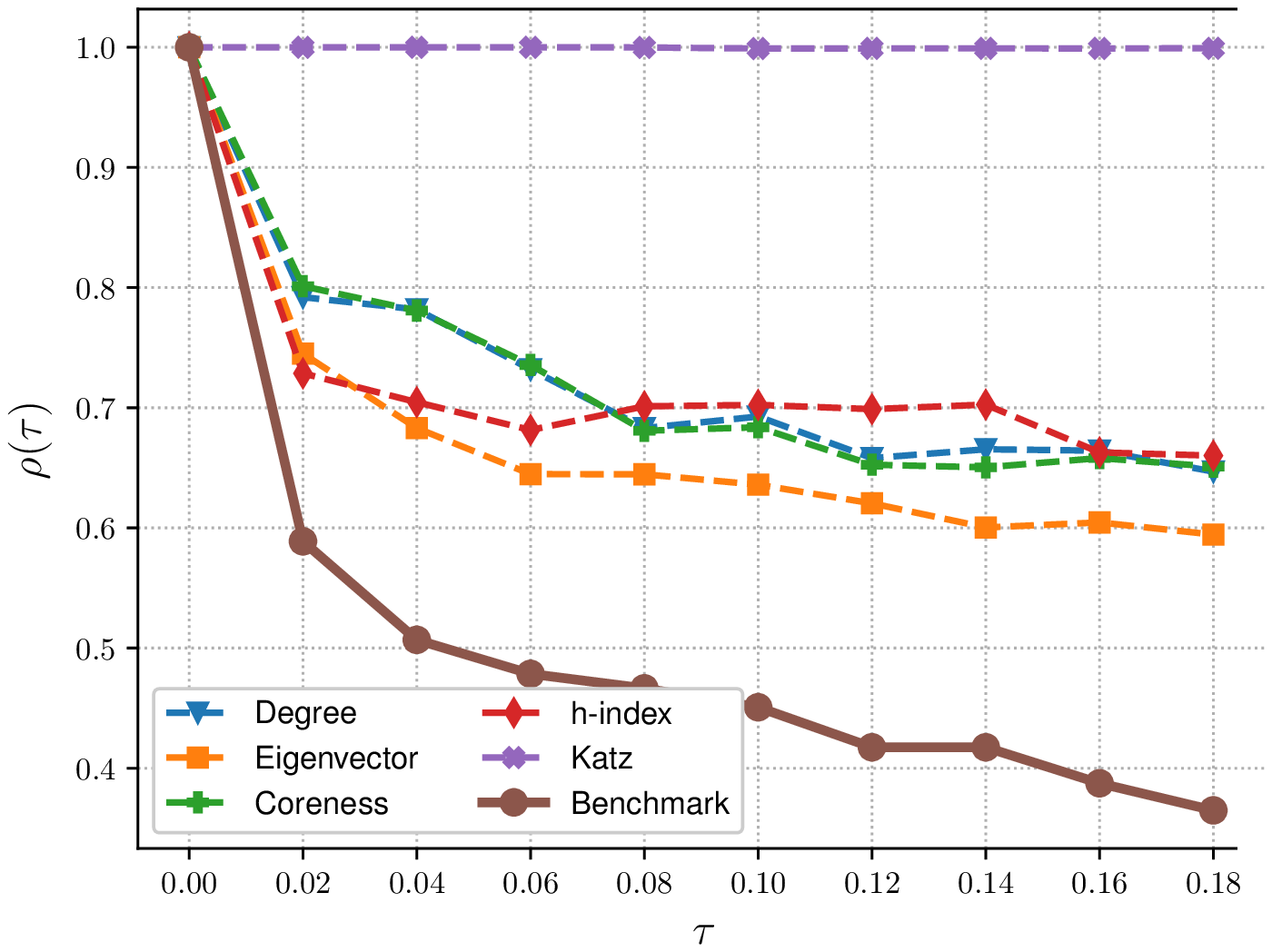}
    \vspace{-2em}
    \caption{BC}
  \end{subfigure}
    \vspace{-0.8em}
     \caption{Effectiveness tests on \textsc{US\_Power\_Grid} dataset: (a-c) Uniform model with $p \in \{0.1,0.3,0.5\}$, (d) BC}
\label{fig:US-POWER_GRID}
\end{figure*}

Figure \ref{fig:US-POWER_GRID} reports the variation of effectiveness for the case of the \textsc{US\_Power\_Grid} dataset as $\tau$ varies.
Considering the Uniform model with $p = 0.1$, the degree has the best effectiveness among all centrality metrics; in fact, it is sufficient to target a fraction $\tau = 0.02$ of nodes to lower the effectiveness from 1 to 0.63. If $p = 0.3$, then the degree and the Eigenvector centrality achieve a comparable effectiveness. In contrast, when $p = 0.5$, or opted for the BC model, the Eigenvector centrality is more effective than the degree. With Katz centrality, the reduction in effectiveness is almost negligible (around $0.01$), for both the Uniform and the BC model. This effect derives from the fact that the most of the nodes in \textsc{US\_Power\_Grid} have degree less than three; thus, the removal of large degree nodes has a devastating impact on effectiveness.

\begin{figure*}[htb]
  \centering  
  \begin{subfigure}{0.245\textwidth}
    \includegraphics [width=\textwidth] {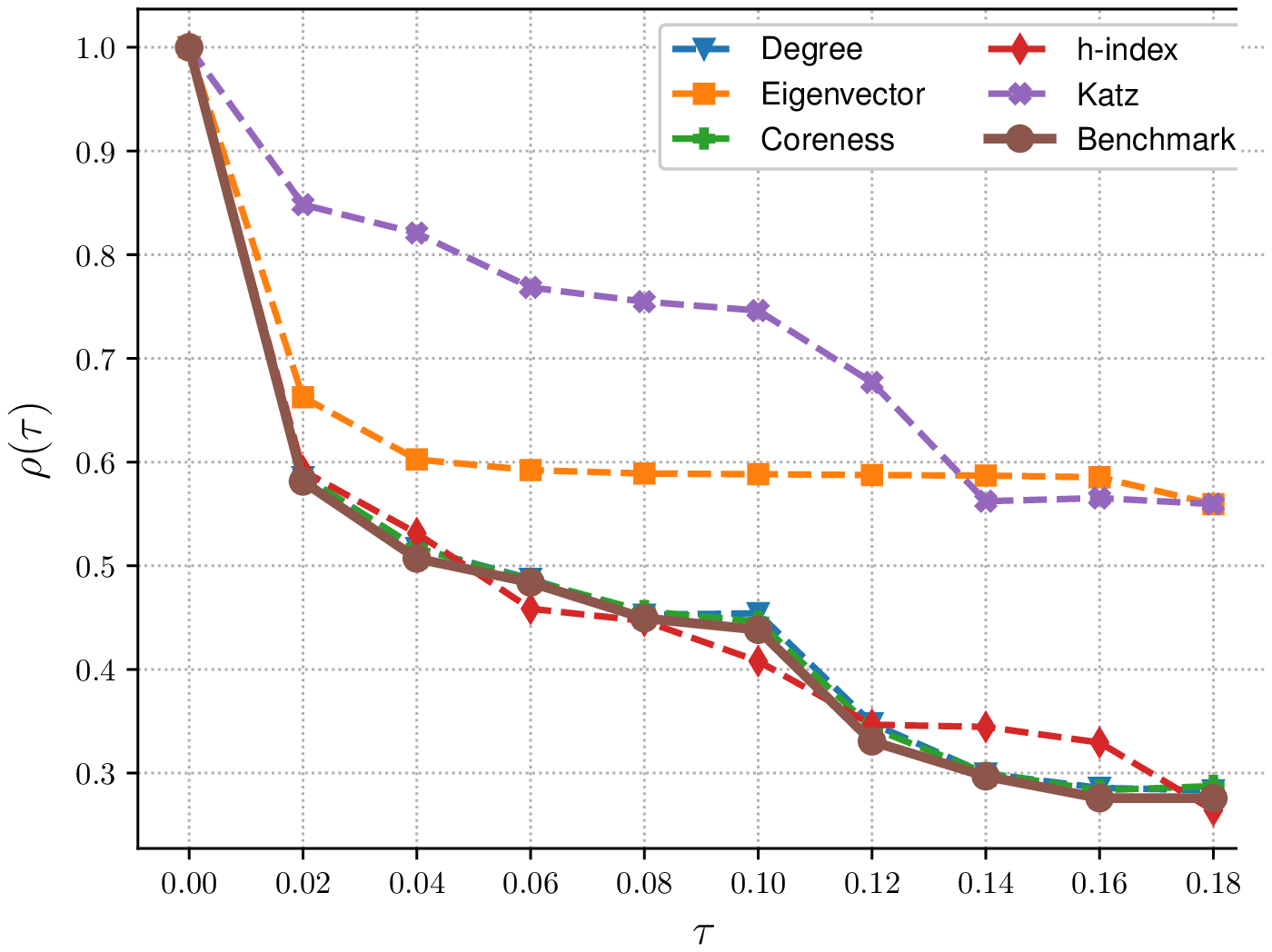}
    \vspace{-2em}
    \caption{Uniform $p=0.1$}
  \end{subfigure}
  \begin{subfigure}{0.245\textwidth}
    \includegraphics [width=\textwidth] {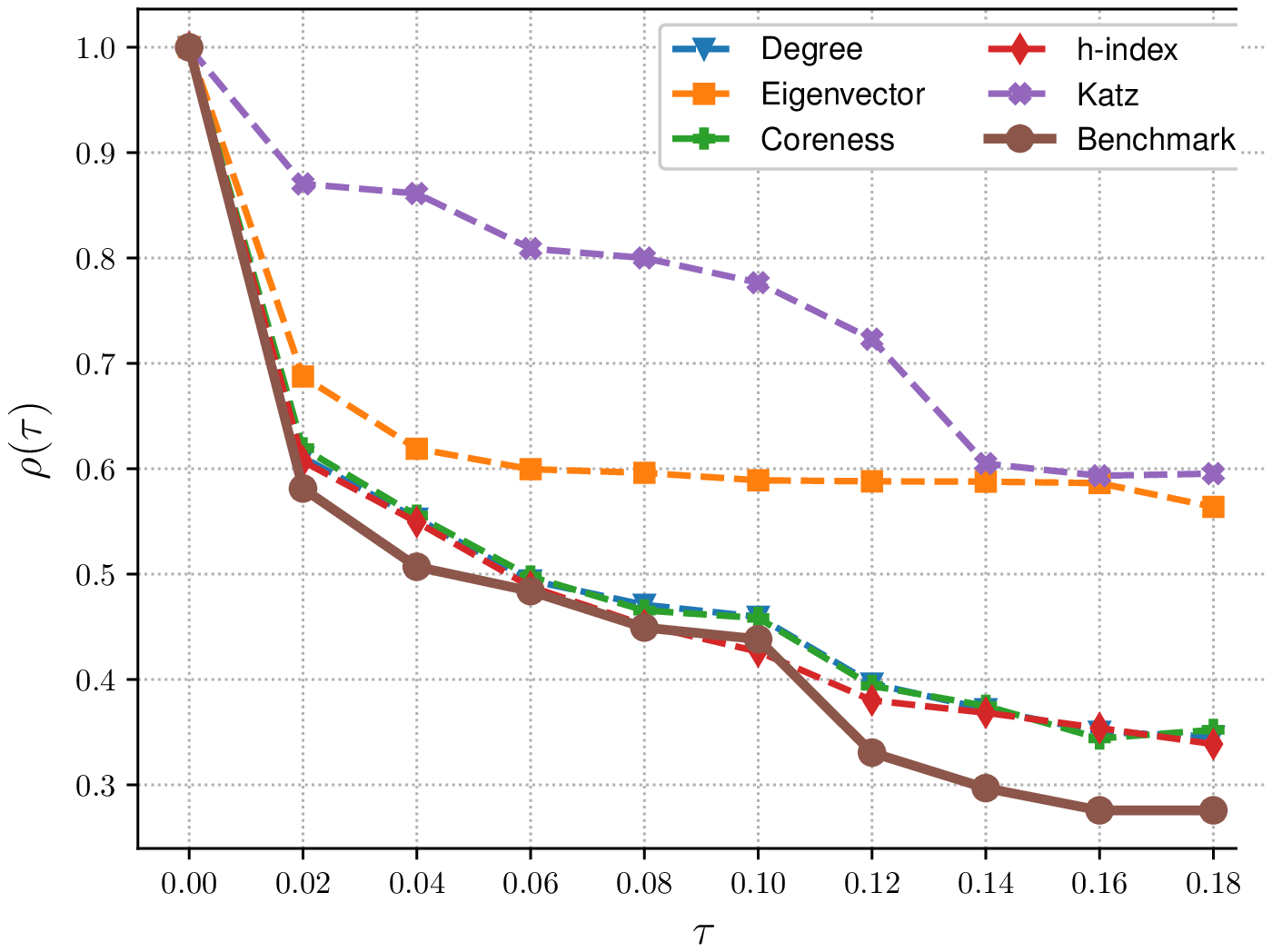}
    \vspace{-2em}
    \caption{Uniform $p=0.3$}
  \end{subfigure}
  \begin{subfigure}{0.245\textwidth}
    \includegraphics [width=\textwidth] {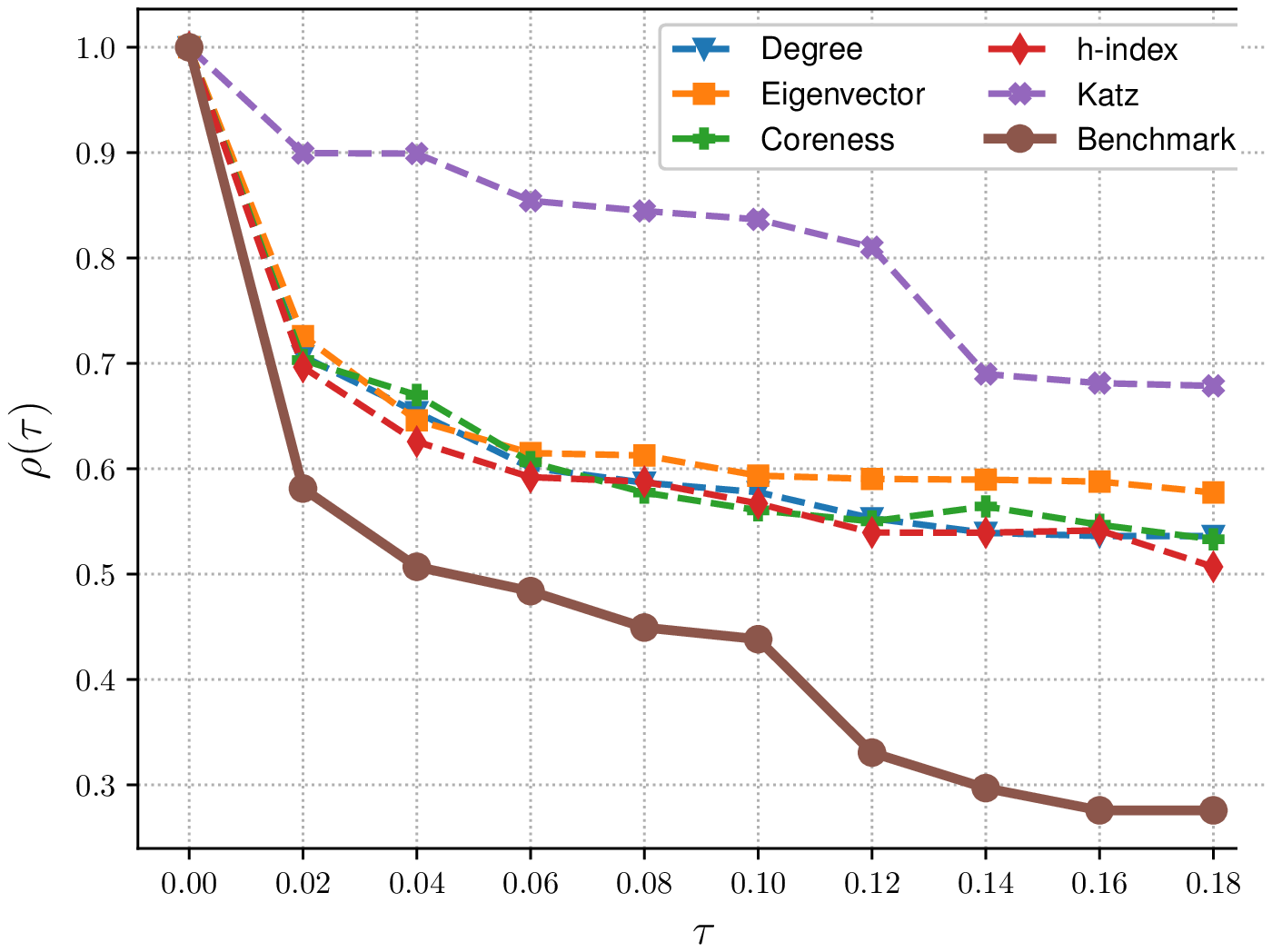}
    \vspace{-2em}
    \caption{Uniform $p=0.5$}
  \end{subfigure}
  \begin{subfigure}{0.245\textwidth}
    \includegraphics [width=\textwidth] {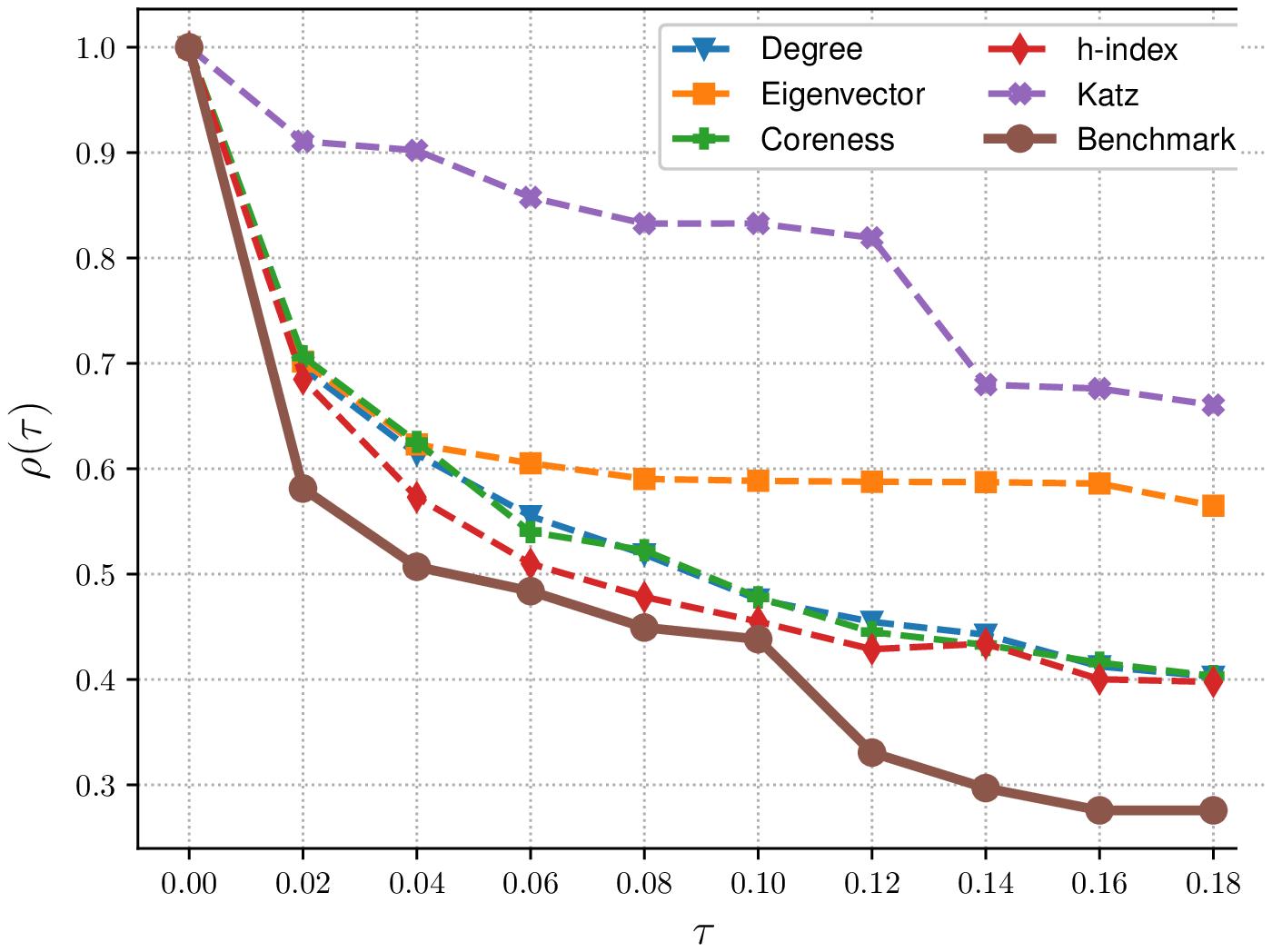}
    \vspace{-2em}
    \caption{BC}
  \end{subfigure}
    \vspace{-0.8em}
     \caption{Effectiveness tests on \textsc{AstroPh} dataset: (a-c) Uniform model with $p \in \{0.1,0.3,0.5\}$, (d) BC.}
\label{fig:AstroPh}
\end{figure*}

As for the \textsc{AstroPh} dataset (Figure \ref{fig:AstroPh}), the degree is more effective than either Eigenvector and Katz centrality.
The effectiveness decay observed in the BC model is more significant than in the Uniform model with $p = 0.3$, independently of the centrality metric adopted. In contrast, in the Uniform model, with $p = 0.5$ emerges a bigger decrease in effectiveness than that observed in the BC model.
Furthermore, the decrease of effectiveness caused by the Katz centrality is smaller than that observed in case of the degree and in the Eigenvector centrality, and it stabilizes for $\tau \geq 0.14$.

Next, we turn our attention to the \textsc{BrightKite} dataset.
In all experimental configurations, the degree is the most effective centrality metrics even though in some cases (e.g., the Uniform model with $p = 0.5$) the gap in effectiveness between the degree and the Eigenvector centrality is almost negligible. If Katz centrality is applied, a slow decrease in effectiveness up to $\tau = 0.16$ is shown. For bigger values of $\tau$, modifications in the graph topology are so significant as to cause a sharp decrease in effectiveness.

The \textsc{BrightKite} dataset is more sensible to node removal than all other datasets considered so far. In the Uniform model with $p = 0.1$, it is sufficient to fix $\tau = 0.02$ to lower effectiveness to 0.24, which is about one third of the value measured in the case of the \textsc{US\_Power\_Grid} dataset.
The reduction in effectiveness in the BC model closely mirrors the Uniform model with $p = 0.3$. 

\textsc{Flickr} is the largest dataset herein investigated with more than two million edges.
It may be viewed as a content network. However, since the process of producing metadata and associating them with images derives from the collaboration of Flickr users, it is predictable that \textsc{Flickr} exhibits some features that make it similar to \textsc{BrightKite}. In fact, the degree generally yields the largest drop in the effectiveness; in addition, Eigenvector centrality usually is more effective than Katz centrality, although this gap is less evident than in other datasets.

Furthermore, effectiveness variations due to degree have been observed to be almost equal to the ones recorded through coreness centrality. This was observed in all datasets herein investigated in both Uniform and BC models. 
This result is consistent with the findings of Lu {\em et al.}~\cite{lu2015hindex}, who highlighted a strong correlation between the degree and the coreness.

The analysis shows that the disruptive power of the h-index (i.e., the rate at which the effectiveness decreases) is comparable to that of the degree and the coreness, provided that the graphs are sufficiently large.

An important exception arises with the \textsc{US-Power-Grid} graph: indeed, most of the nodes exhibit an h-index of either one or two; thus, the process of choosing nodes on the basis of their h-index has a minor impact on the effectiveness.

\subsection{Coverage of Centrality Measures}
\label{sub:coverage}
\begin{figure*}[htb]
  \centering  
  \begin{subfigure}{0.245\textwidth}
    \includegraphics [width=\textwidth] {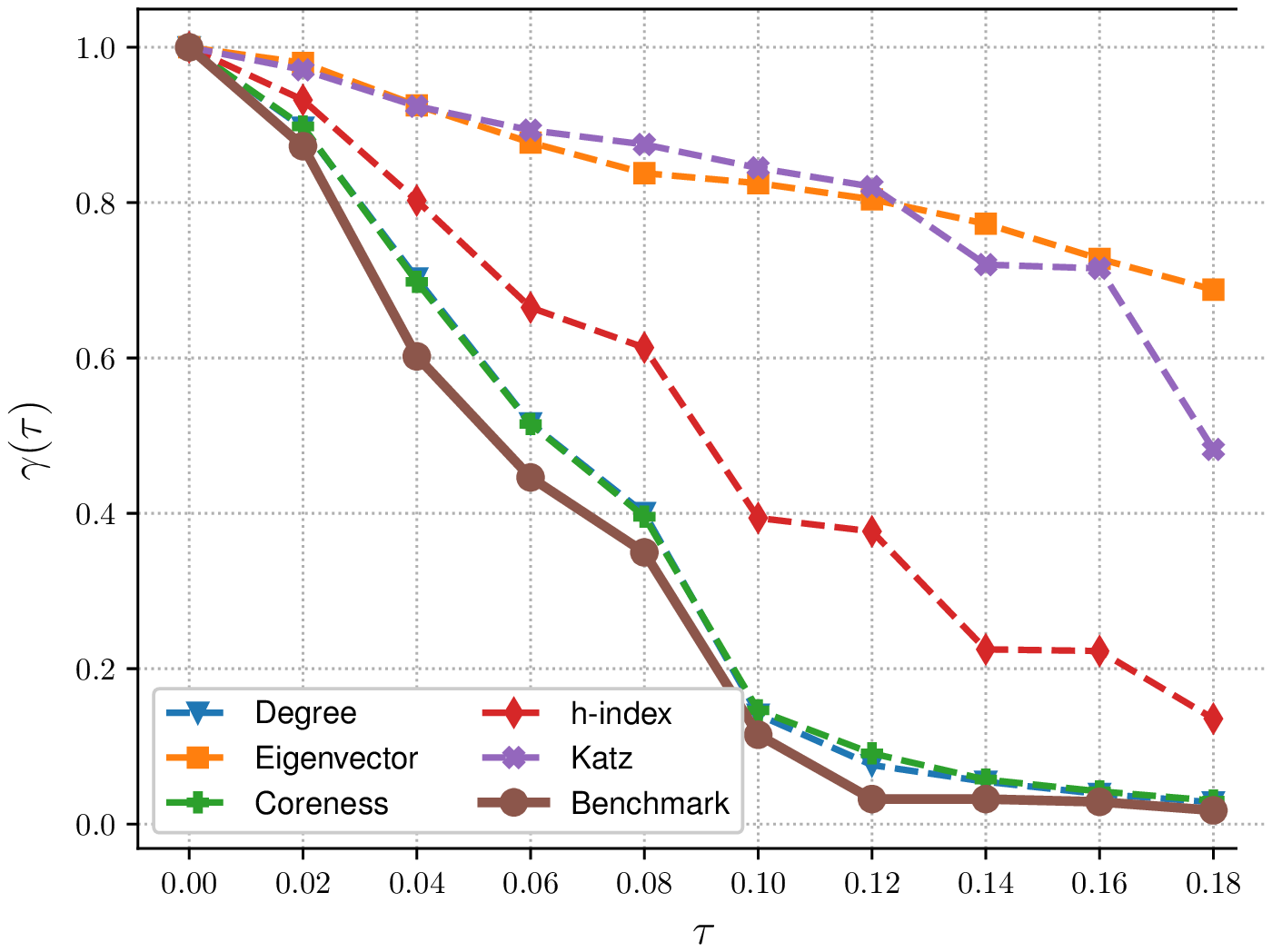}
     \vspace{-2em}
    \caption{Uniform $p=0.1$}
  \end{subfigure}
  \begin{subfigure}{0.245\textwidth}
    \includegraphics [width=\textwidth] {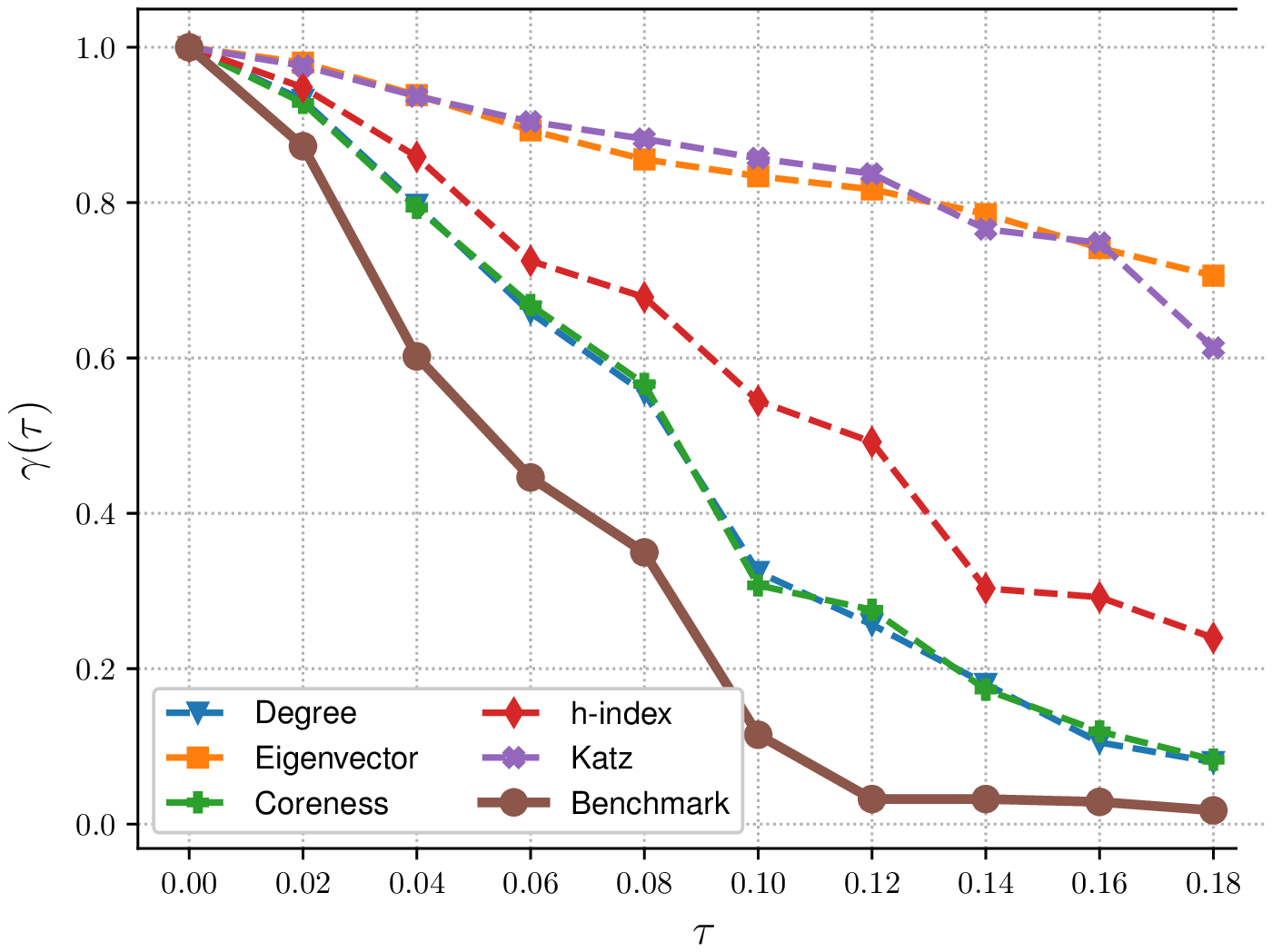}
     \vspace{-2em}
    \caption{Uniform $p=0.3$}
  \end{subfigure}
  \begin{subfigure}{0.245\textwidth}
    \includegraphics [width=\textwidth] {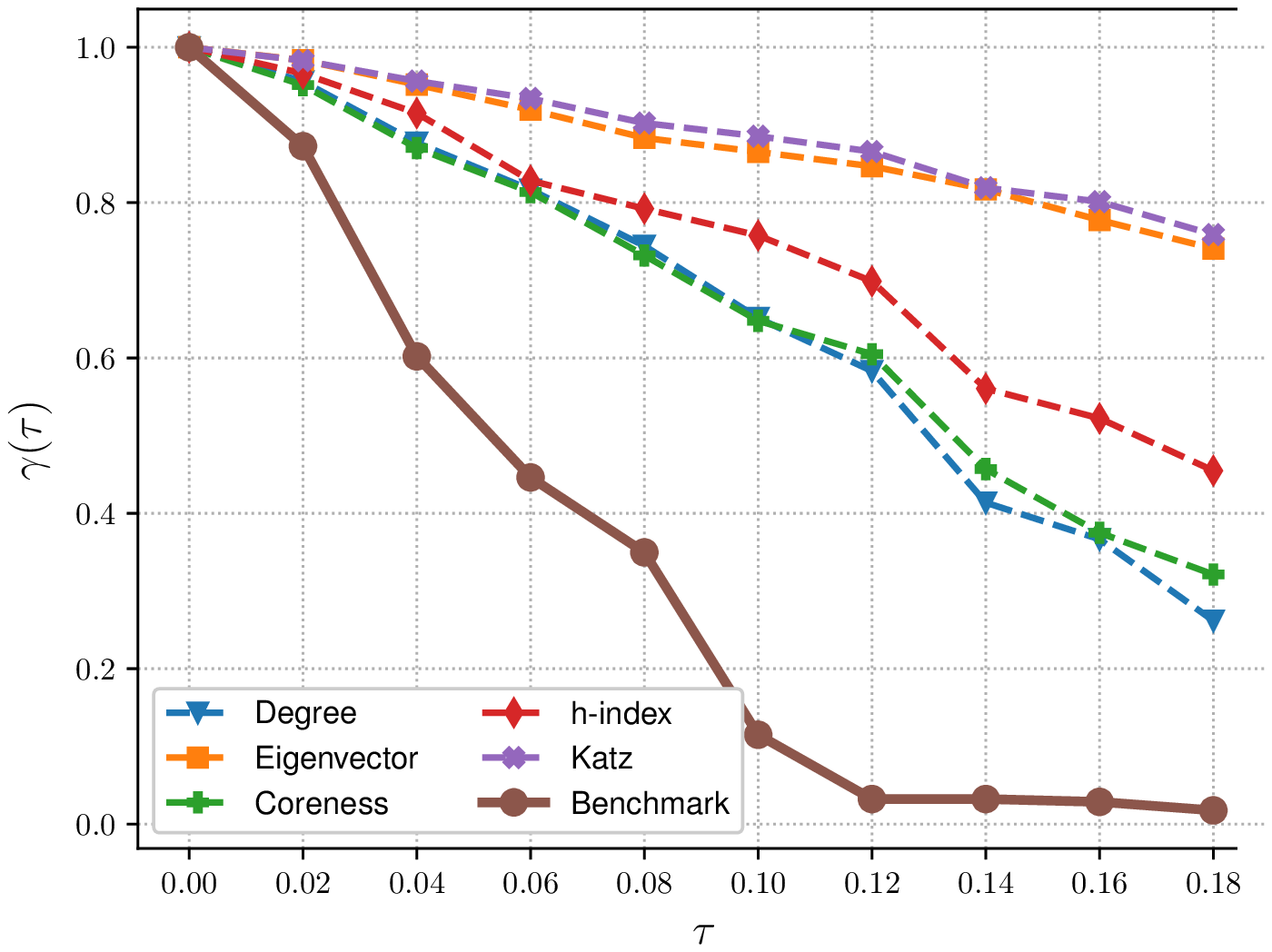}
     \vspace{-2em}
    \caption{Uniform $p=0.5$}
  \end{subfigure}
  \begin{subfigure}{0.245\textwidth}
    \includegraphics [width=\textwidth] {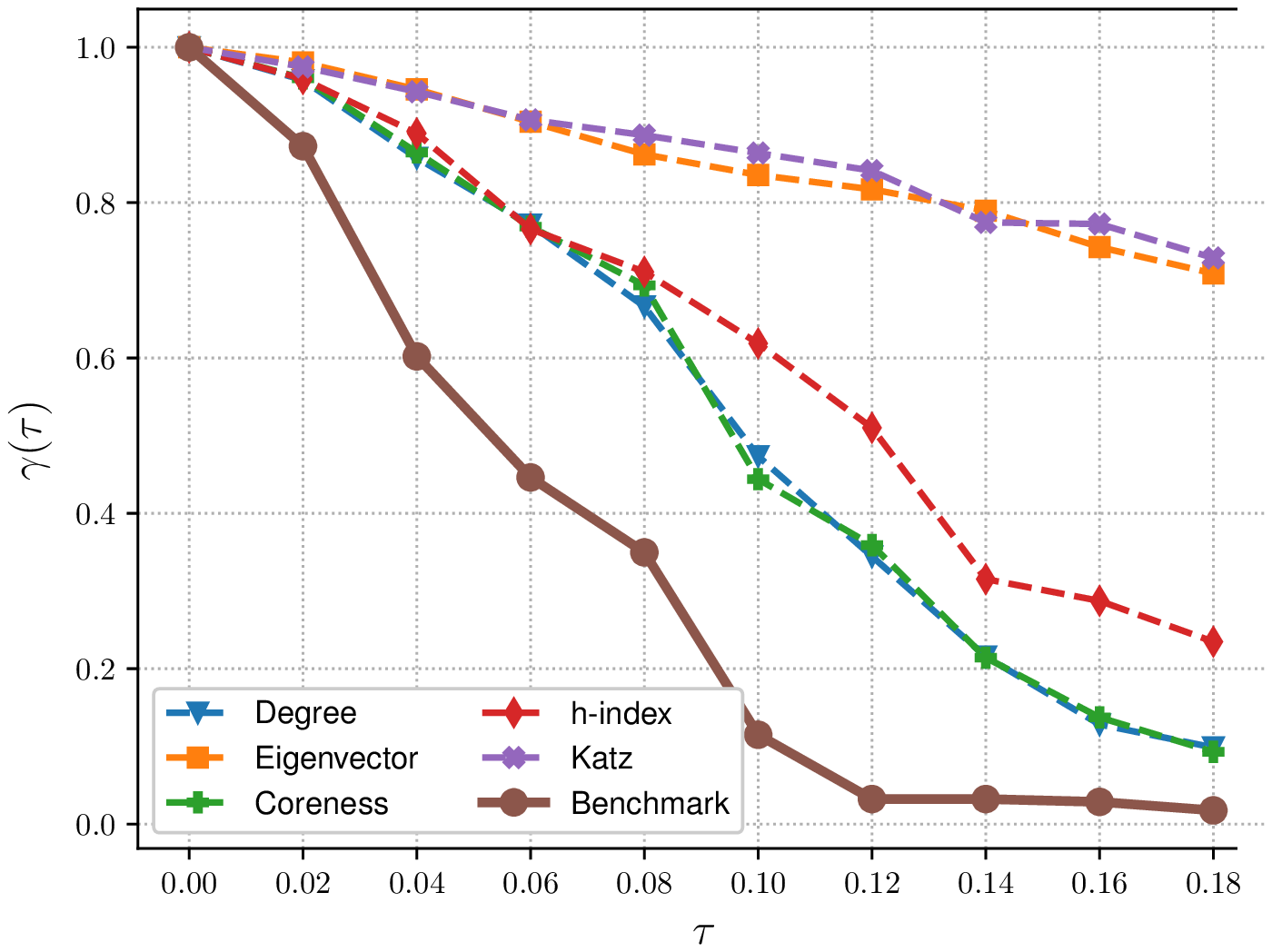}
    \vspace{-2em}
    \caption{BC}
  \end{subfigure}
  \vspace{-0.8em}
     \caption{Coverage tests on \textsc{US-Power-Grid} dataset: (a-c) Uniform model with $p \in \{0.1,0.3,0.5\}$, (d) BC.}
\label{fig:US-POWER_GRID-cc}
\end{figure*}

\begin{figure*}[htb]
  \centering  
  \begin{subfigure}{0.245\textwidth}
    \includegraphics [width=\textwidth] {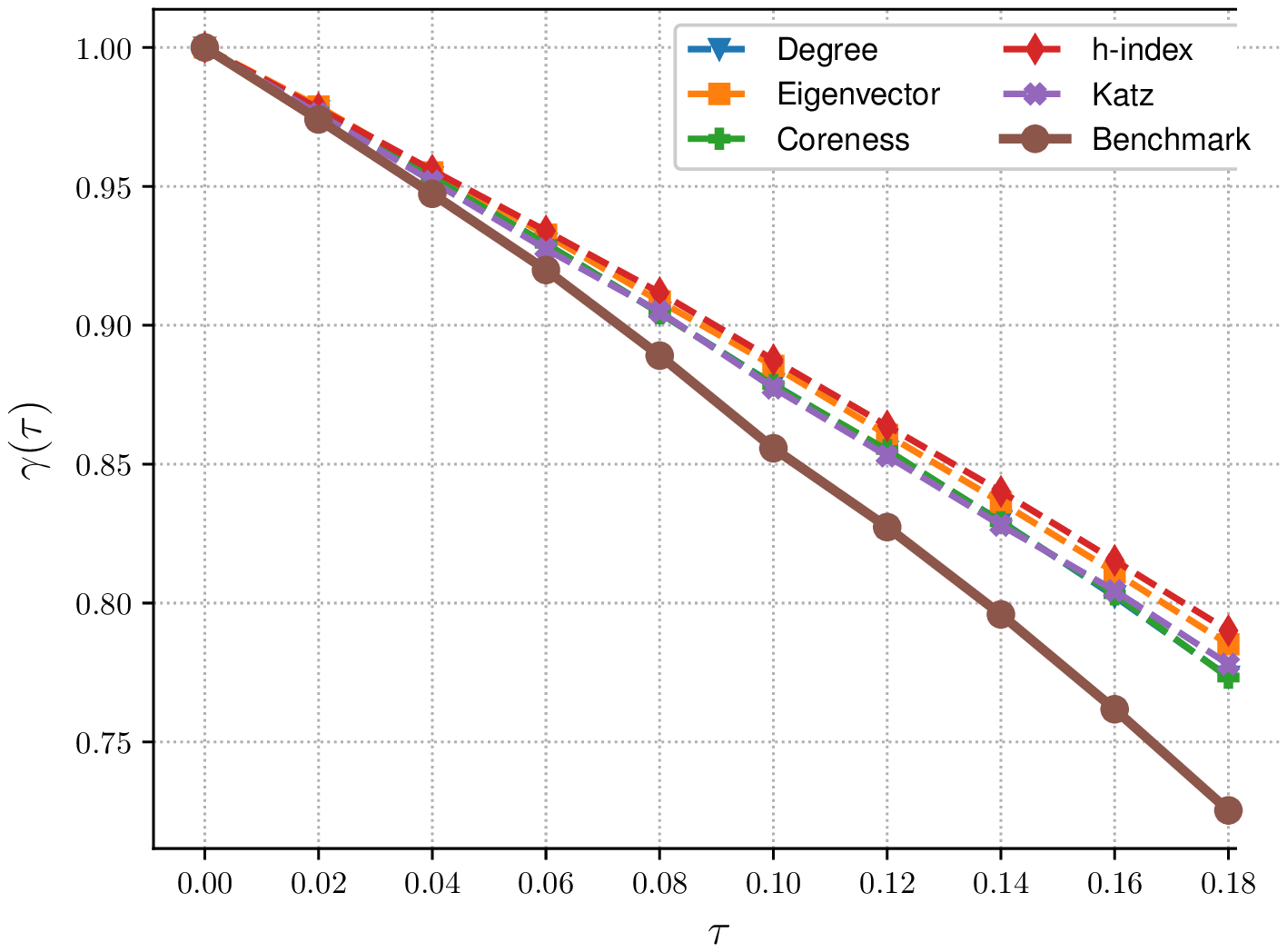}
    \vspace{-2em}
    \caption{Uniform $p=0.1$}
  \end{subfigure}
  \begin{subfigure}{0.245\textwidth}
    \includegraphics [width=\textwidth] {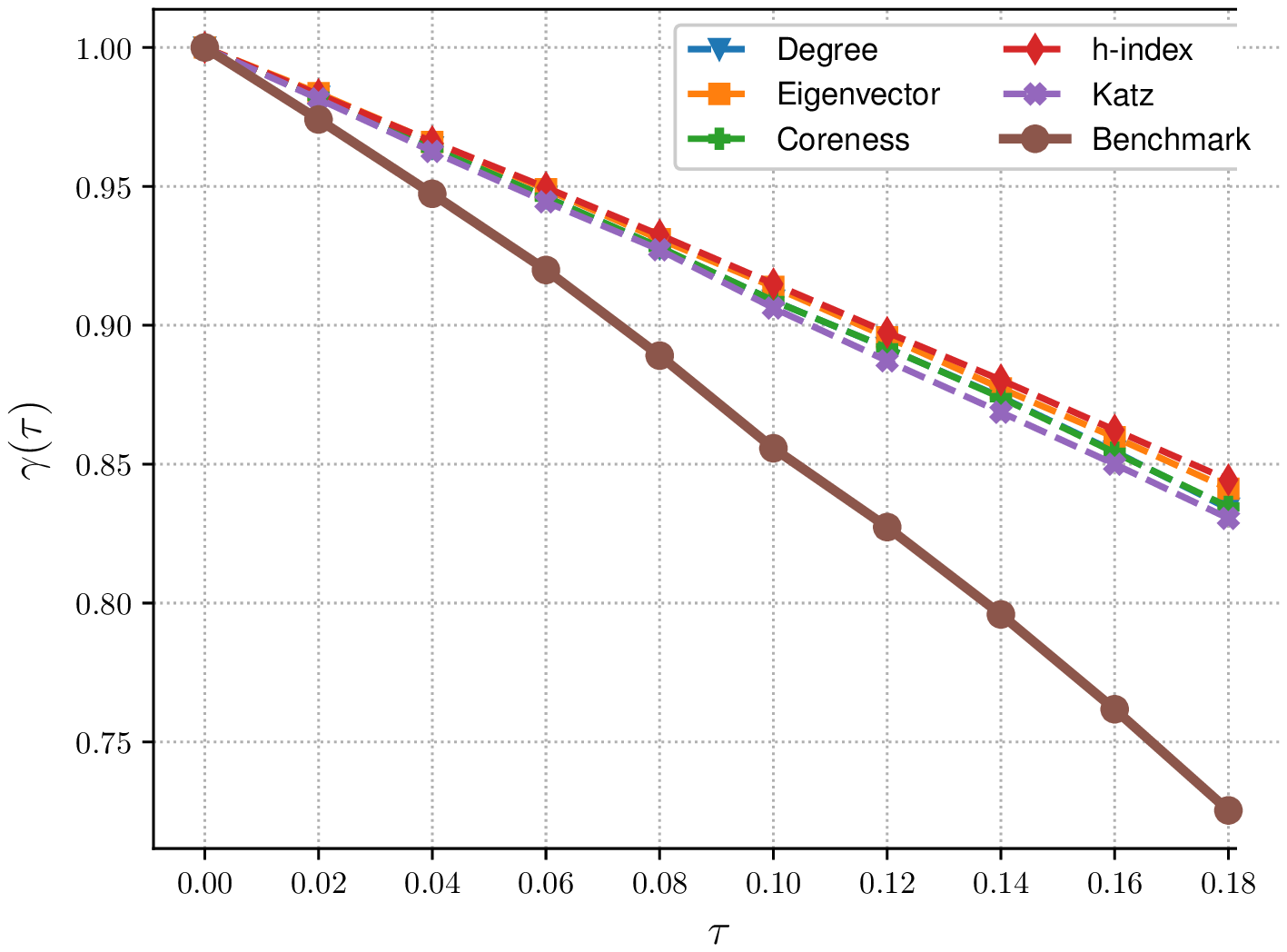}
    \vspace{-2em}
    \caption{Uniform $p=0.3$}
  \end{subfigure}
  \begin{subfigure}{0.245\textwidth}
    \includegraphics [width=\textwidth] {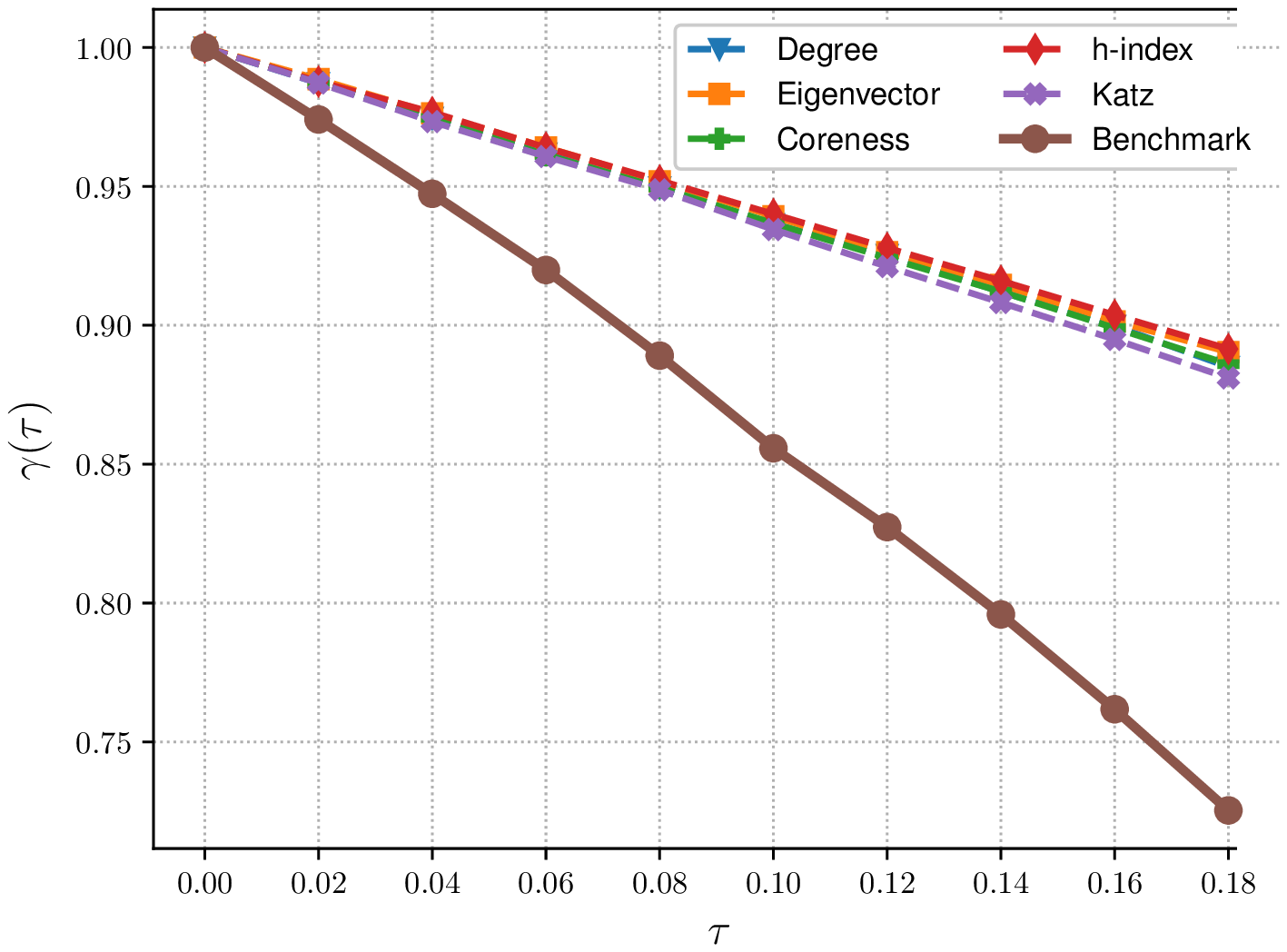}
    \vspace{-2em}
    \caption{Uniform $p=0.5$}
  \end{subfigure}
  \begin{subfigure}{0.245\textwidth}
    \includegraphics [width=\textwidth] {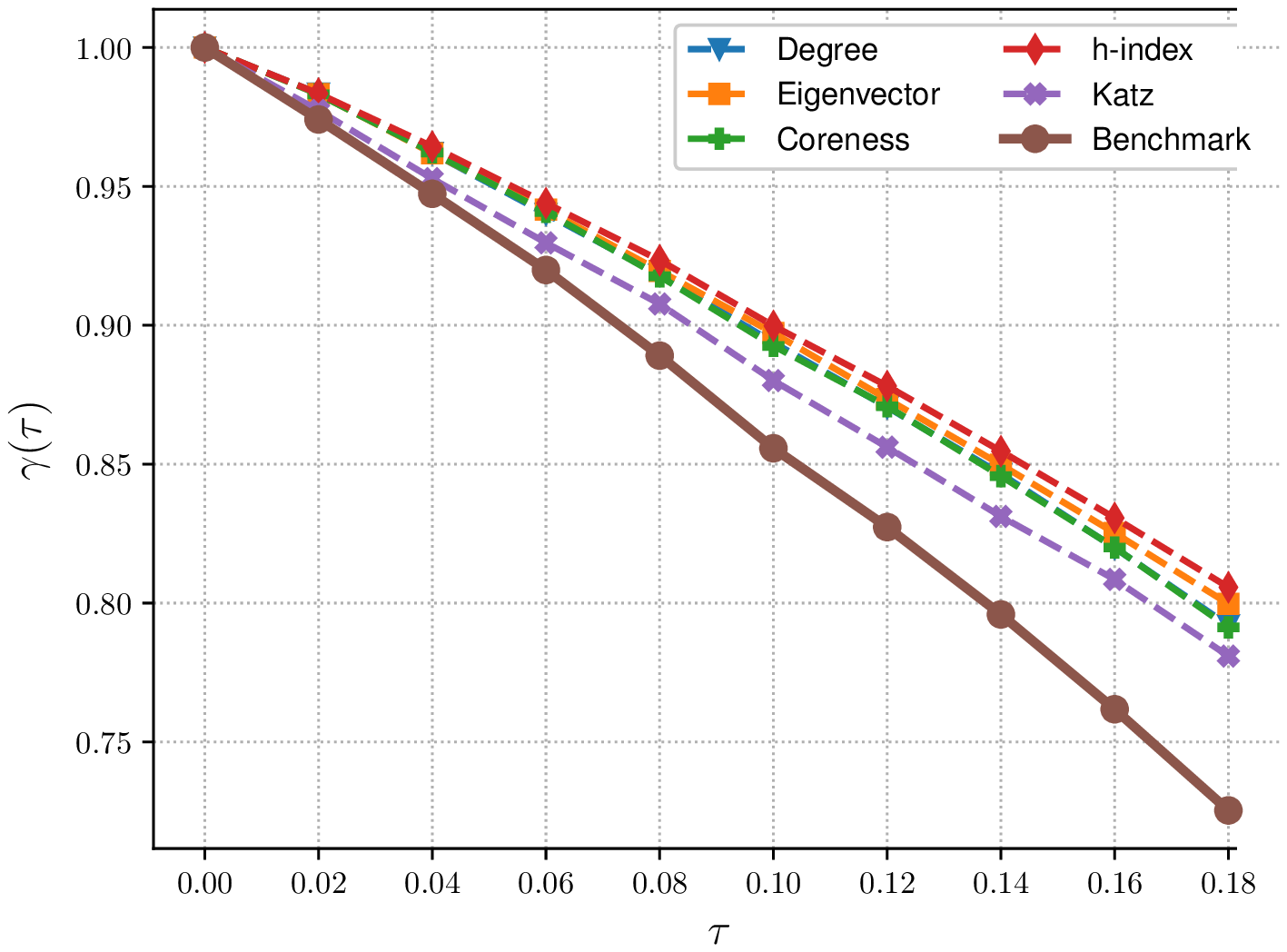}
    \vspace{-2em}
    \caption{BC}
  \end{subfigure}
  \vspace{-0.8em}
     \caption{Coverage tests on \textsc{AstroPh} dataset: (a-c) Uniform model with $p \in \{0.1,0.3,0.5\}$, (d) BC.}
\label{fig:AstroPh-cc}
\end{figure*}

\begin{figure*}[htb]
  \centering
  \begin{subfigure}{0.245\textwidth}
    \includegraphics [width=\textwidth] {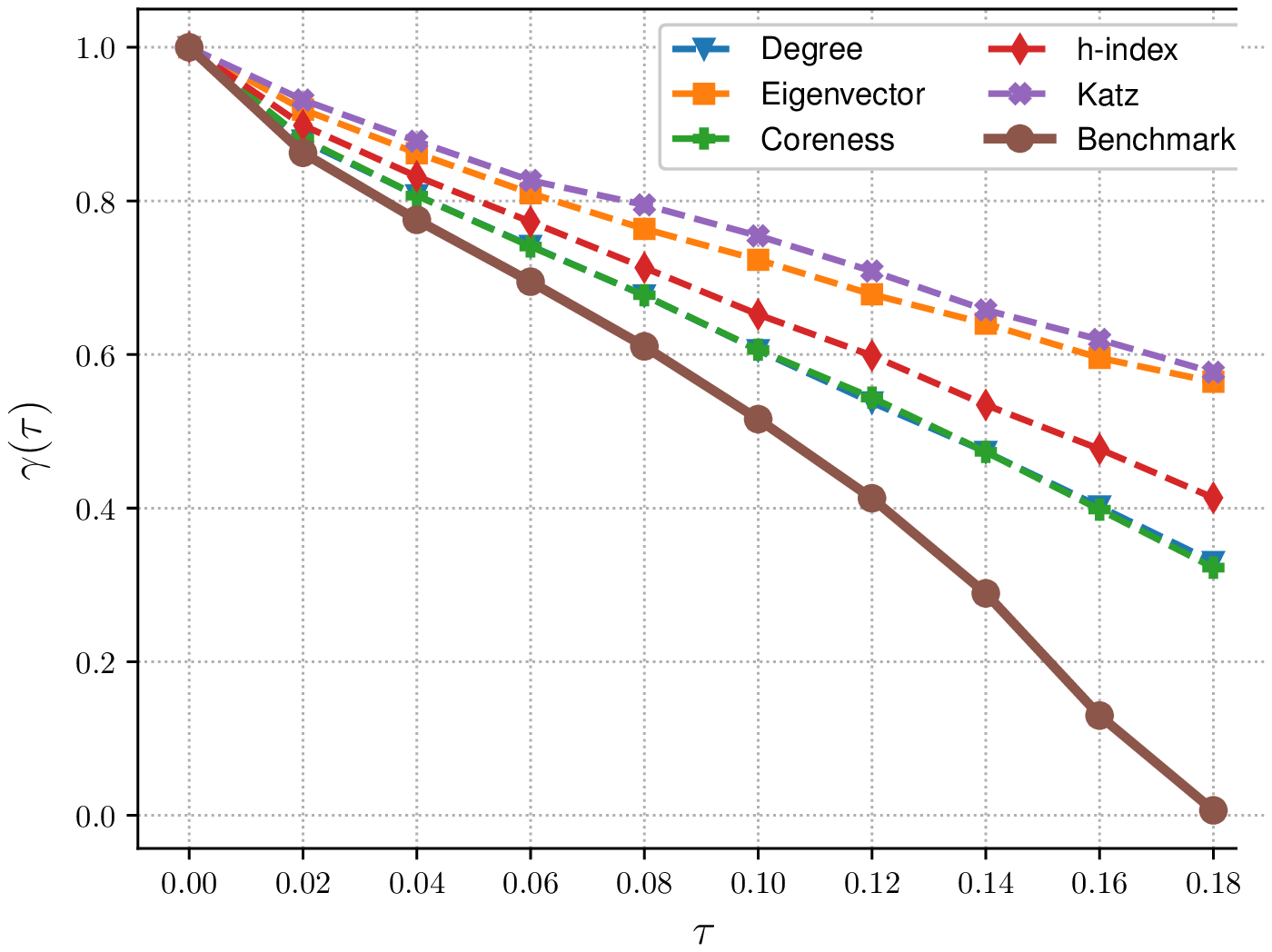}
    \vspace{-2em}
    \caption{Uniform $p=0.1$}
  \end{subfigure}
  \begin{subfigure}{0.245\textwidth}
    \includegraphics [width=\textwidth] {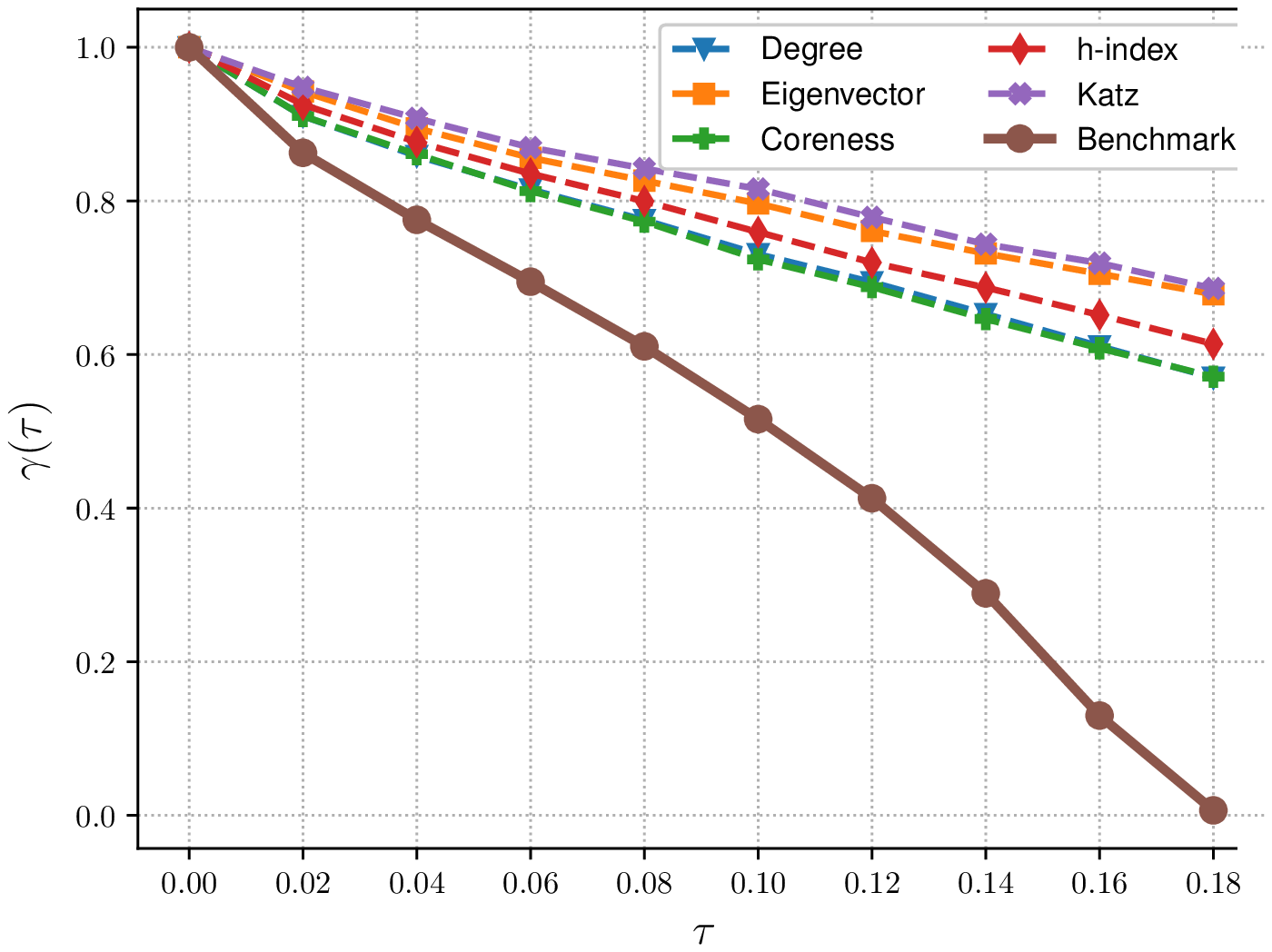}
    \vspace{-2em}
    \caption{Uniform $p=0.3$}
  \end{subfigure}
  \begin{subfigure}{0.245\textwidth}
    \includegraphics [width=\textwidth] {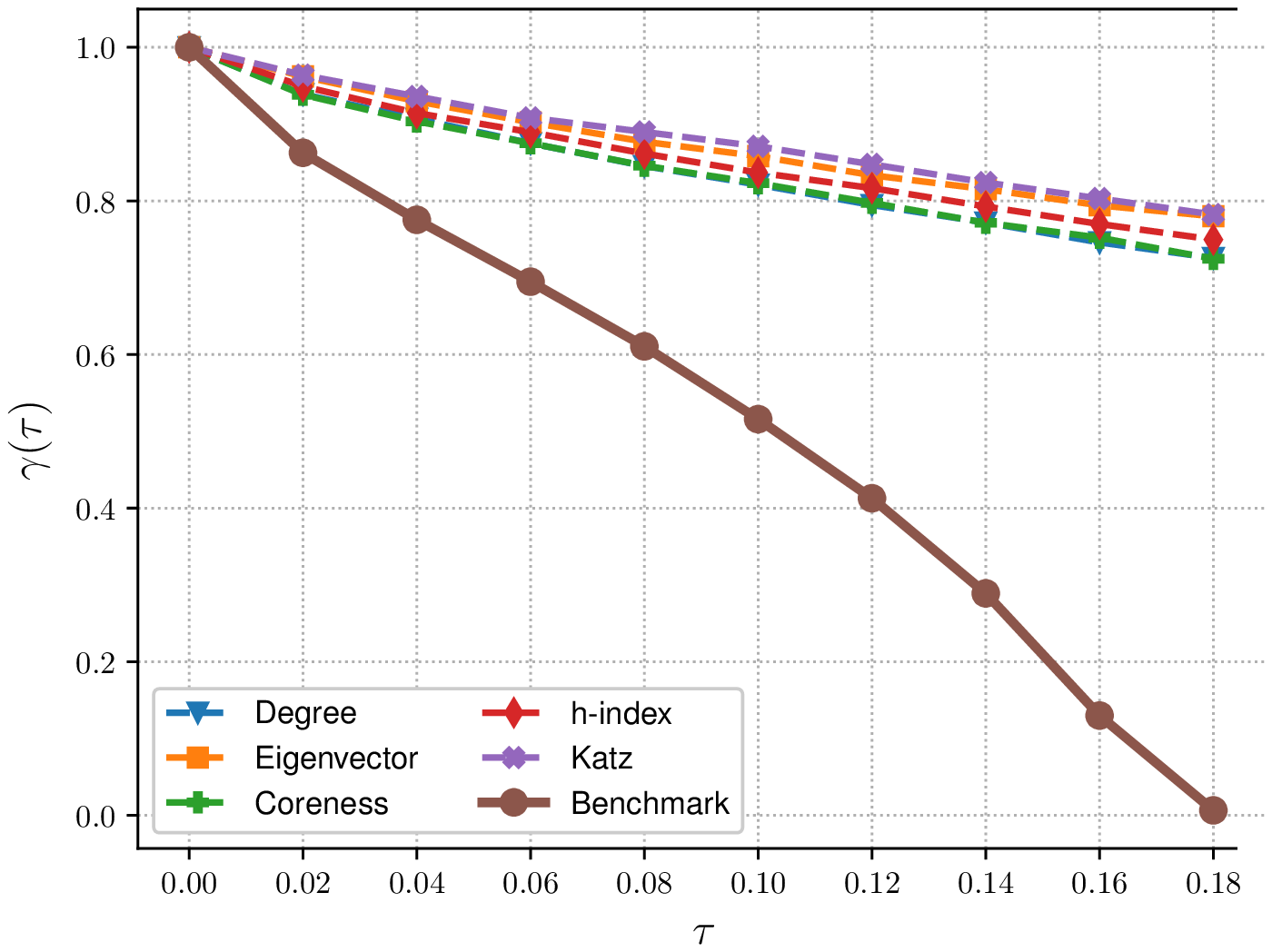}
    \vspace{-2em}
    \caption{Uniform $p=0.5$}
  \end{subfigure}
  \begin{subfigure}{0.245\textwidth}
    \includegraphics [width=\textwidth] {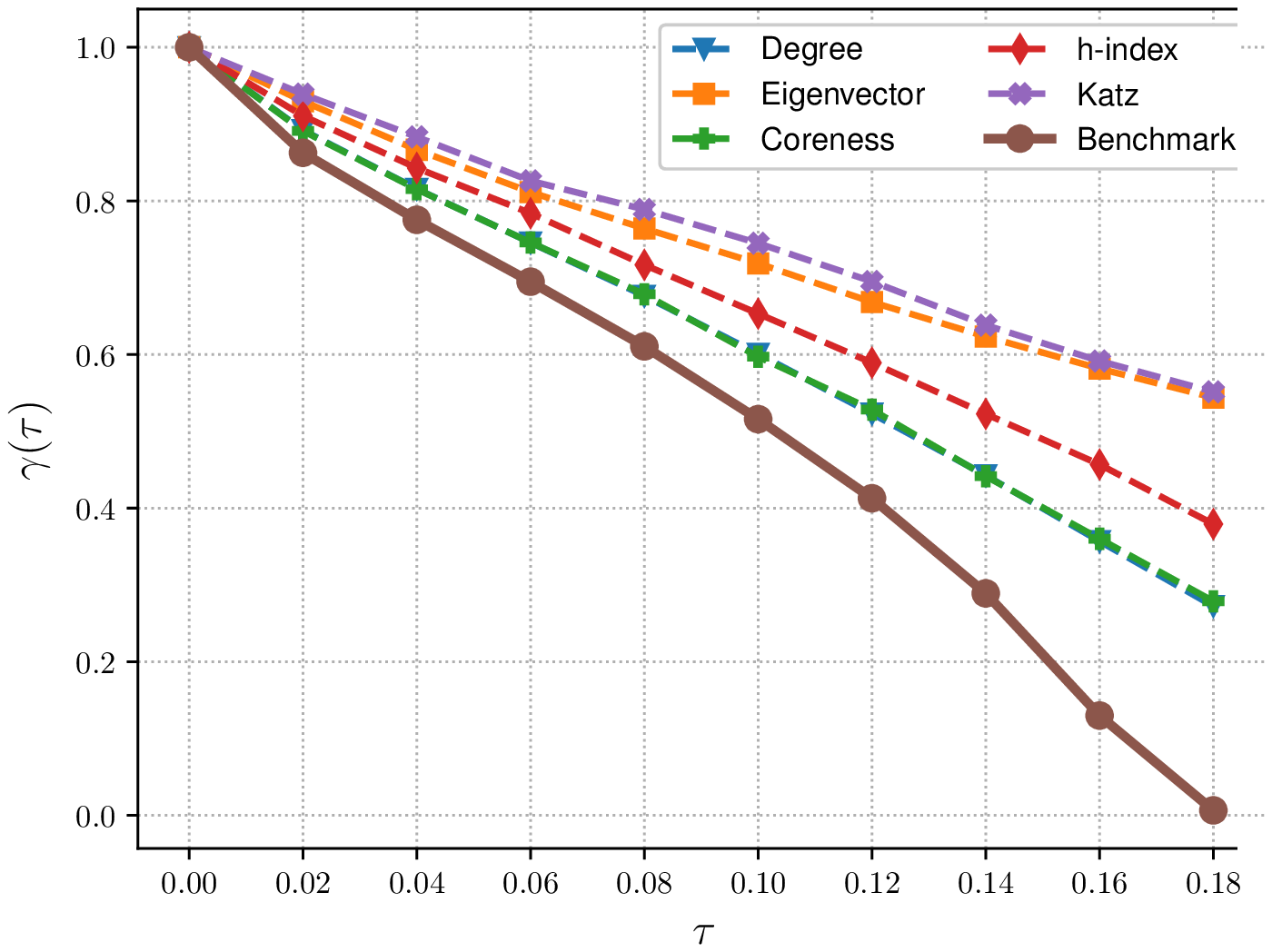}
    \vspace{-2em}
    \caption{BC}
  \end{subfigure}
  \vspace{-0.8em}
      \caption{Coverage tests on \textsc{Brightkite} dataset: (a-c) Uniform model with $p \in \{0.1,0.3,0.5\}$, (d) BC.}
  \label{fig:Brightkite-cc}
\end{figure*}

\begin{figure*}[htb]
  \centering  
  \begin{subfigure}{0.245\textwidth}
    \includegraphics [width=\textwidth] {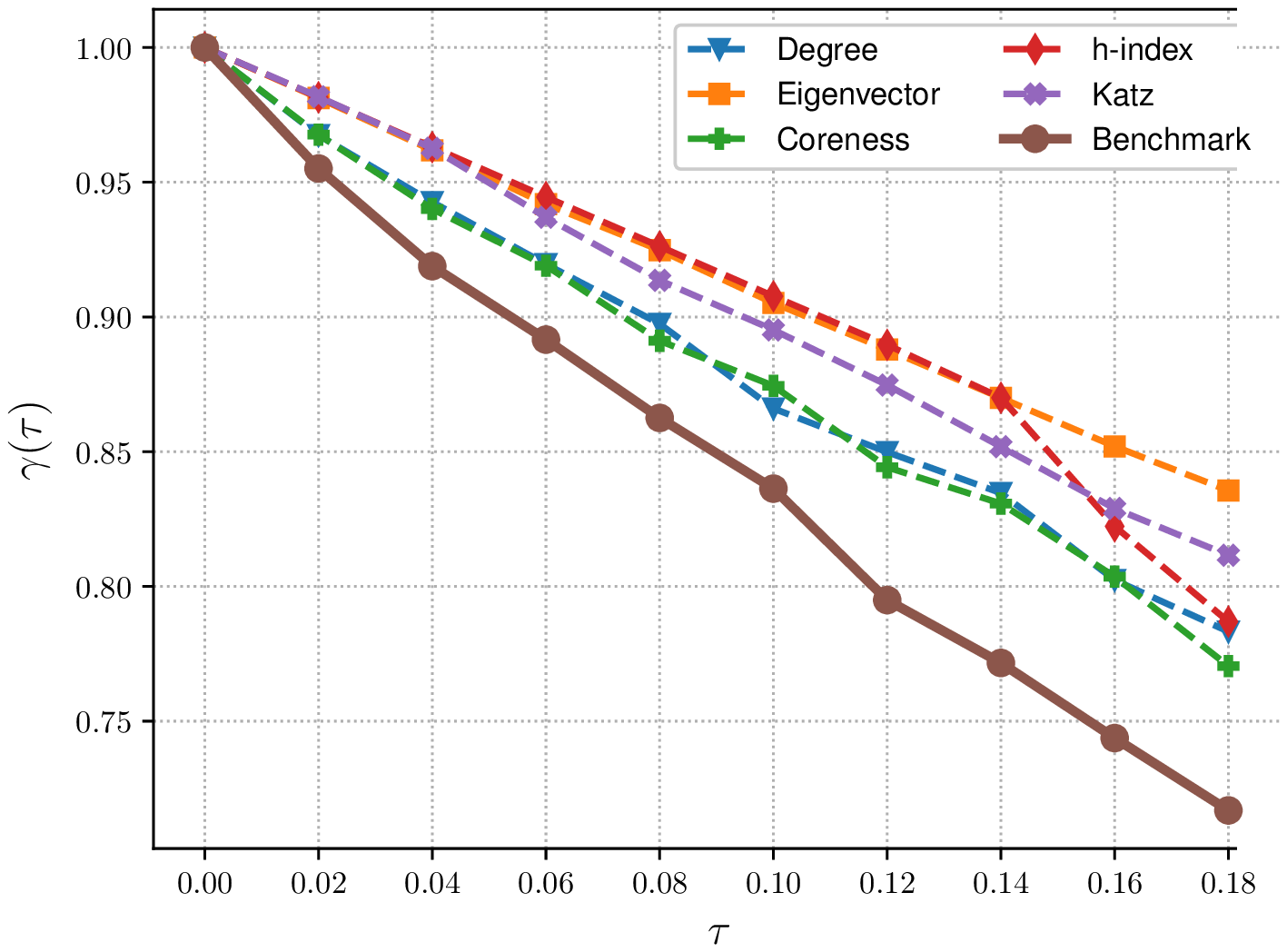}
    \vspace{-2em}
    \caption{Uniform $p=0.1$}
  \end{subfigure}
  \begin{subfigure}{0.245\textwidth}
    \includegraphics [width=\textwidth] {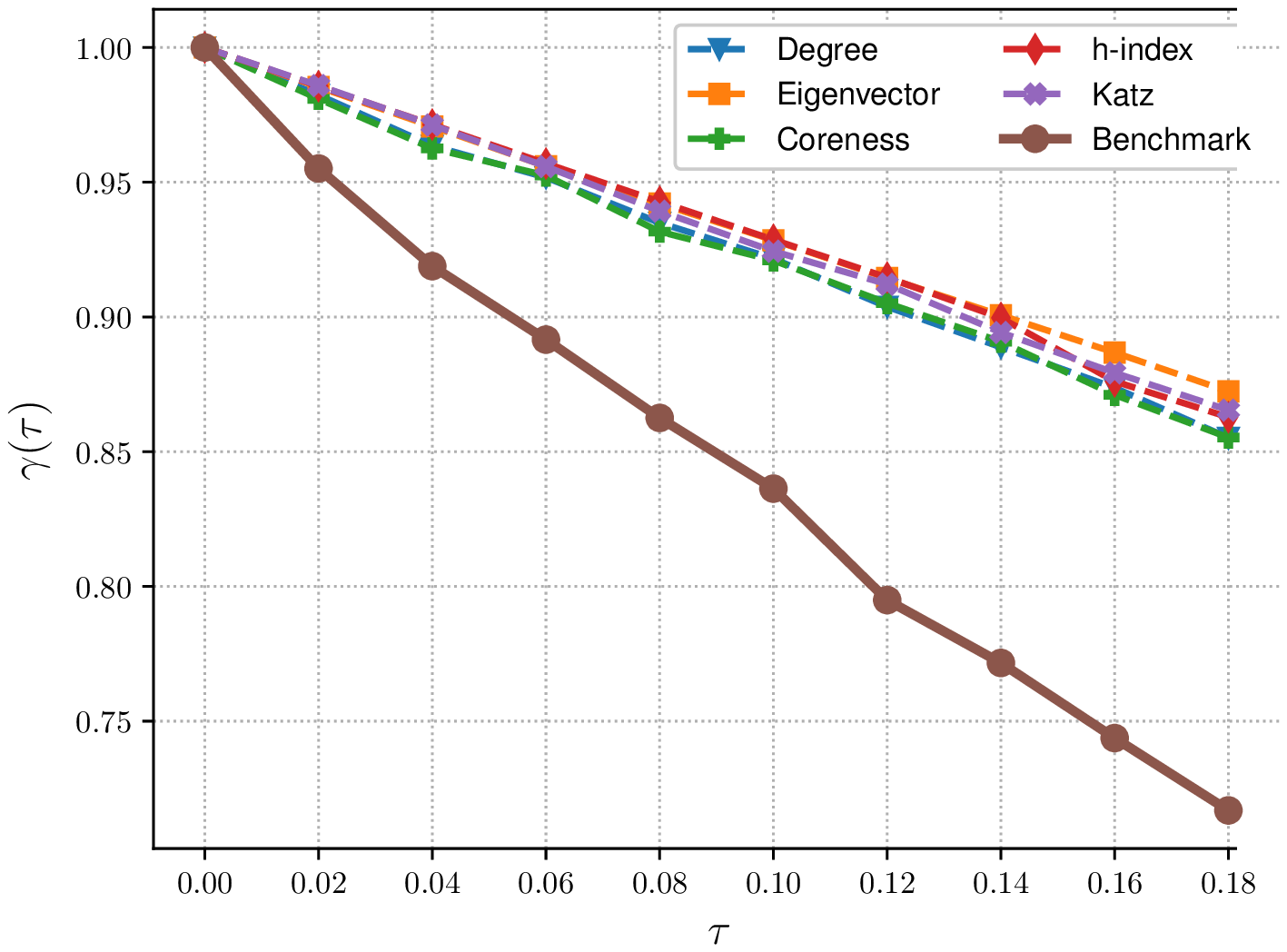}
    \vspace{-2em}
    \caption{Uniform $p=0.3$}
  \end{subfigure}
  \begin{subfigure}{0.245\textwidth}
    \includegraphics [width=\textwidth] {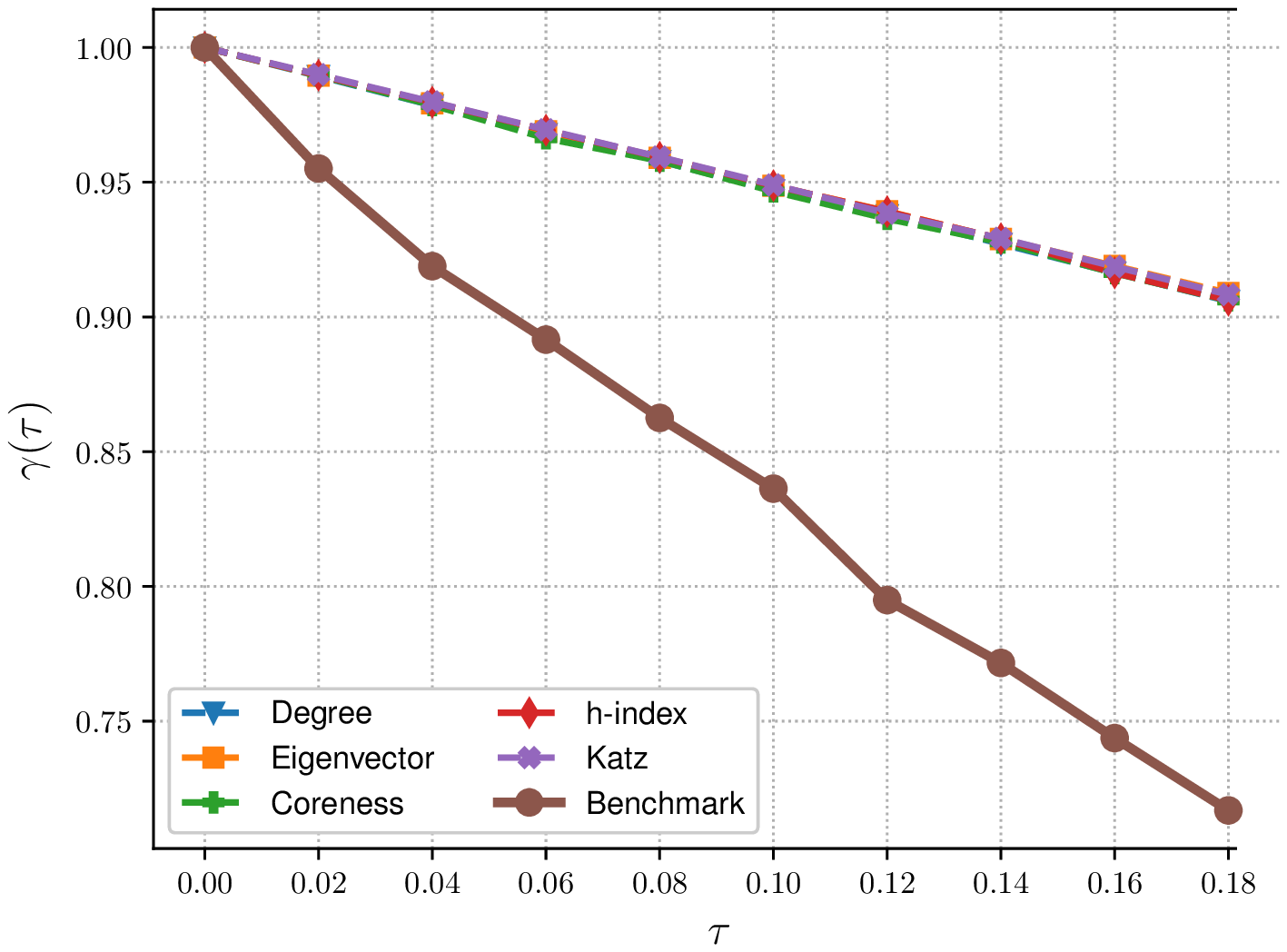}
    \vspace{-2em}
    \caption{Uniform $p=0.5$}
  \end{subfigure}
  \begin{subfigure}{0.245\textwidth}
    \includegraphics [width=\textwidth] {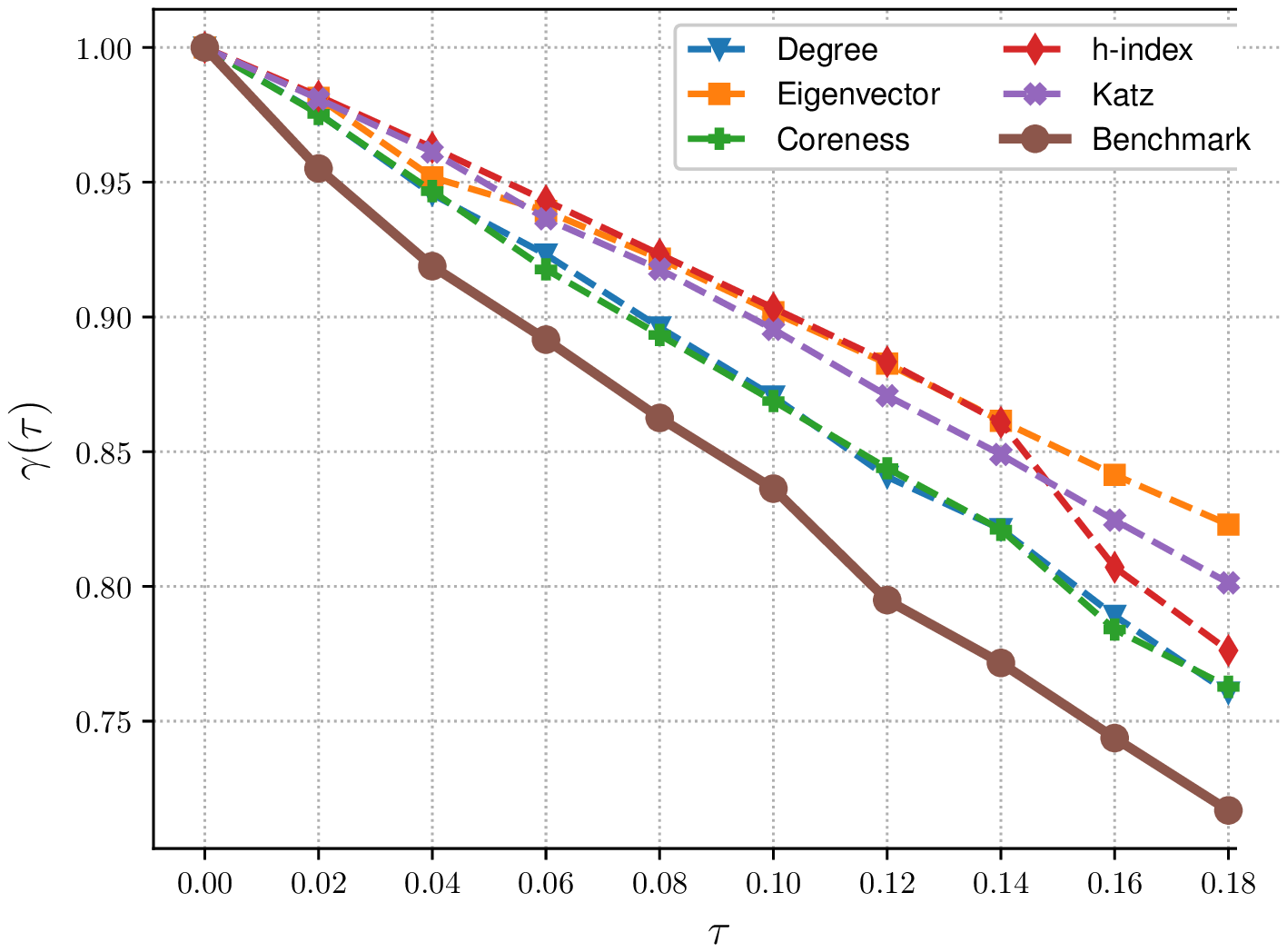}
    \vspace{-2em}
    \caption{BC}
  \end{subfigure}
    \vspace{-0.8em}
     \caption{Coverage tests on \textsc{\textsc{Flickr}} dataset: (a-c) Uniform model with $p \in \{0.1,0.3,0.5\}$, (d) BC.}
\label{fig:Flickr-cc}
\end{figure*}

Figures \ref{fig:US-POWER_GRID-cc}-\ref{fig:Flickr-cc} report the variation of coverage as function of the fraction $\tau$ of removed nodes. 

In the \textsc{US\_Power\_Grid} dataset (Fig.~\ref{fig:US-POWER_GRID-cc}), the degree and coreness always yield the largest reduction in coverage. Herein, is needed to target a relatively large fraction of nodes (i.e., $\tau \geq 0.14$) to observe a sensible reduction of coverage.
Such a trend is likely dependent on the topological structure of power grid networks~\cite{albert2004structural}: these kind of networks, in fact, display a high redundancy level; thus, they may endure the failure of a relatively small number of nodes before becoming disconnected.

In the \textsc{AstroPh}, \textsc{BrightKite} and \textsc{Flickr} datasets (Figures \ref{fig:AstroPh-cc}, \ref{fig:Brightkite-cc}, and \ref{fig:Flickr-cc}, respectively), the coverage decreases in an almost linear fashion, independently of the model adopted to encode node failure probability and the centrality metric selected to target nodes. Once again, the degree and the coreness are responsible for the largest reduction in coverage, although in some configurations (e.g., the \textsc{Flickr} dataset in the Uniform model with $p = 0.5$), all centrality metrics cause the same amount of reduction in terms of coverage.

In the \textsc{AstroPh} dataset (Fig.~\ref{fig:AstroPh-cc}), the coverage dropped from 1 to 0.77 (in both the BC and Uniform model with $p = 0.1$); analogously, the observed reduction in coverage in \textsc{Flickr} (Fig.~\ref{fig:Flickr-cc}) was from 1 to 0.74 (in both the BC and Uniform model with $p = 0.1$).

By contrast, in the \textsc{Brightkite} dataset (Fig.~\ref{fig:Brightkite-cc}) a stronger decrease in coverage emerged rather than in the \textsc{AstroPh} and \textsc{Flickr} datasets. This especially if the Uniform model with $p = 0.1$, or the BC model are considered.
Remember that the \textsc{Brightkite} dataset edges identify friendship relationships among members. Therefore, there are few individuals who accumulate a large fraction of friendship relationships, and whose removal implies a quick fragmentation of the \textsc{Brightkite} graph into isolated components.

\subsection{A comparison with NetShield}
\label{sub:comparison-netshield}

To complete the analysis, this section compares the centrality metrics herein used and the NetShield algorithm~\cite{chen2016node} (Sect.~\ref{sub:spectral-graph-role}).

As previously asserted, the NetShield algorithm takes an undirected graph $G$ and an integer $k$ as input and returns a set of $k$ nodes $\mathcal{S}^{NS}(k)$ to remove from $G$ to achieve the biggest drop in $\lambda_1$.

To perform experimental analysis, we had to slightly modify the evaluation protocol illustrated above. In fact, the worst-case time complexity of NetShield amounts to $O(nk^2 + m)$, being $n$ and $m$ the number of nodes and edges of $G$: if, as in our previous tests, we assumed that $k$ is of the same order of magnitude of $n$, then the worst-case time complexity of NetShield would be cubic in $n$. This make the approach unfeasible on even moderately large graphs.
Therefore, we consider values of $k$ ranging from $1$ to $15$.
Due to space limitations, we report our results only in case of degree (which generally proved to be the most effective centrality metric among those we considered). To keep our notation consistent, we call $\mathcal{S}^{d}(k)$ the set of $k$ nodes selected by the degree.

We introduced the parameter $\beta$ to quantify the relative effectiveness of the NetShield algorithm against the degree:

\begin{equation}
\label{eqn:beta1}
  \beta = \frac{\tilde{\lambda}_1(NS)}{\tilde{\lambda}_1(d)}
\end{equation}

Herein, $\tilde{\lambda}_1(NS)$ (resp., $\tilde{\lambda}_1(d)$) is the spectral radius of $G$ after deleting nodes in $\mathcal{S}^{NS}(k)$ (resp., $\mathcal{S}^{d}(k)$).

\begin{table}
\centering
    \begin{tabular}{ccccc}
     \hline \hline 
    $k$ &  \textsc{{US\_Power\_Grid}} &  \textsc{AstroPh} &  \textsc{BrightKite} &  \textsc{Flickr} \\
    \hline
    1  &        0.960 &    0.972 &       0.956 &   0.984 \\
    2  &        0.990 &    0.991 &       0.914 &   0.979 \\
    5  &        1.153 &    1.052 &       1.073 &   0.971 \\
    10 &        1.403 &    1.063 &       1.258 &   0.984 \\
    15 &        1.633 &    1.174 &       1.516 &   1.031 \\
    \hline
    \end{tabular}
\caption{Values of $\beta$ as $k$ increases in the Uniform model with $p = 0.1$.} 
~\label{tbl:lambda-ratio-uniform-01}
\end{table}

Analogously, we introduced the parameter $\gamma$ to quantify the relative coverage of the NetShield algorithm against the degree:

\begin{equation}
\label{eqn:beta2}
    \gamma = \frac{\tilde{c}(NS)}{\tilde{c}(d)}
\end{equation}

Herein, $\tilde{c}(NS)$ (resp., $\tilde{c}(d)$) is the LCC size of $G$ after deleting nodes in $\mathcal{S}^{NS}(k)$ (resp., $\mathcal{S}^{d}(k)$).
Observe that if $\beta < 1$ (resp. $\beta > 1$), then the NetShield algorithm is more (resp., less) effective than the degree.
Analogously, if $\gamma < 1$ (resp. $\gamma > 1$), then the NetShield algorithm has a better (resp., worse) coverage (resp., less) effective than the degree.

The values of $\beta$ in the Uniform and BC models as $k$ increases are reported in Tables \ref{tbl:lambda-ratio-uniform-01}-\ref{tbl:lambda-ratio-best-connected} (i.e., $p=0.1$ in Table~\ref{tbl:lambda-ratio-uniform-01}, $p=0.3$ in Table~\ref{tbl:lambda-ratio-uniform-03}, $p=0.5$ in Table~\ref{tbl:lambda-ratio-uniform-05}, and BC in Table~\ref{tbl:lambda-ratio-best-connected}).

\begin{table}
\centering
     \begin{tabular}{ccccc}
     \hline \hline 
    $k$ &  \textsc{{US\_Power\_Grid}} &  \textsc{AstroPh} &  \textsc{BrightKite} &  \textsc{Flickr} \\
   \hline
      1  &        0.933 &    0.991 &       0.958 &   0.996 \\
      2  &        0.967 &    0.983 &       0.937 &   0.993 \\
      5  &        1.071 &    1.000 &       0.888 &   0.987 \\
      10 &        1.140 &    1.029 &       0.944 &   0.976 \\
      15 &        1.199 &    1.057 &       1.056 &   0.970 \\
    \hline
    \end{tabular}
\caption{Values of $\beta$ as $k$ increases in the Uniform model with $p = 0.3$.} 
~\label{tbl:lambda-ratio-uniform-03}
\end{table}

In both the Uniform and BC model, the $\beta$ parameter usually ranges from 0.88 to 0.99 if $k$ values less than (or equal to) five are chosen. In concrete scenarios (e.g., if the need is to block the spread of an epidemics in a human community), the largest reduction in the spectral radius by blocking the least number of nodes is required; thus, NetShield is the best weapon in our arsenal even if it can be time-consuming. The degree effectiveness is quite close to that of NetShield if $k < 10$, but the calculation of the degree is much faster than the application of the NetShield algorithm; thus, the degree can be considered as a valid alternative to NetShield on large graphs. 

\begin{table}
\centering
    \begin{tabular}{ccccc}
     \hline \hline 
    $k$ &  \textsc{{US\_Power\_Grid}} &  \textsc{AstroPh} &  \textsc{BrightKite} &  \textsc{Flickr} \\
    \hline
      1  &        0.916 &    0.995 &       0.972 &   0.997 \\
      2  &        0.920 &    0.991 &       0.945 &   0.995 \\
      5  &        0.956 &    0.995 &       0.907 &   0.991 \\
      10 &        1.086 &    1.006 &       0.877 &   0.983 \\
      15 &        1.120 &    1.029 &       0.939 &   0.978 \\
    \hline
    \end{tabular}
\caption{Values of $\beta$ as $k$ increases in the Uniform model with $p = 0.5$.} 
~\label{tbl:lambda-ratio-uniform-05}
\end{table}

In contrast, if $k \geq 10$, the degree is more effective than NetShield, with an improvement up to 60\%.

\begin{table}
\centering
    \begin{tabular}{ccccc}
     \hline \hline 
    $k$ &  \textsc{{US\_Power\_Grid}} &  \textsc{AstroPh} &  \textsc{BrightKite} &  \textsc{Flickr} \\
    \hline
     1 &        0.926 &    0.998 &       0.979 &   0.999 \\
     2 &        0.919 &    1.000 &       0.984 &   0.999 \\
     5 &        0.979 &    0.997 &       0.966 &   1.000 \\
    10 &        0.995 &    1.010 &       0.969 &   0.996 \\
    15 &        1.056 &    1.020 &       0.933 &   0.994 \\
    \hline
    \end{tabular}
\caption{Values of $\beta$ as $k$ increases in the BC model.} 
~\label{tbl:lambda-ratio-best-connected}
\end{table}

Let us now consider the relative coverage in Tables \ref{tbl:lcc-ratio-uniform-01}-\ref{tbl:lcc-ratio-best-connected} (i.e., $p=0.1$ in Table~\ref{tbl:lcc-ratio-uniform-01}, $p=0.3$ in Table~\ref{tbl:lcc-ratio-uniform-03}, $p=0.5$ in Table~\ref{tbl:lcc-ratio-uniform-05}, and BC in Table~\ref{tbl:lcc-ratio-best-connected}).

\begin{table}
\centering
    \begin{tabular}{ccccc}
     \hline \hline 
    $k$ &  \textsc{{US\_Power\_Grid}} &  \textsc{AstroPh} &  \textsc{BrightKite} &  \textsc{Flickr} \\
    \hline
  1 &        0.978 &    1.004 &       1.069 &   1.001 \\
  2 &        1.139 &    1.008 &       1.072 &   1.001 \\
  5 &        1.396 &    1.030 &       1.080 &   1.001 \\
 10 &        1.754 &    1.174 &       1.758 &   1.002 \\
 15 &        1.867 &    1.234 &       1.700 &   1.001 \\
   \hline
    \end{tabular}
\caption{Values of $\gamma$ as $k$ increases in the Uniform model with $p = 0.1$.} 
~\label{tbl:lcc-ratio-uniform-01}
\end{table}

In the Uniform model, the degree significantly outperforms NetShield on the \textsc{{US\_Power\_Grid}} dataset, regardless of the $p$ value. The $\gamma$ value growth is proportional to $k$, which indicates the superiority of the degree in reducing the LCC size compared to NetShield. 

\begin{table}
\centering
    \begin{tabular}{ccccc}
     \hline \hline 
    $k$ &  \textsc{{US\_Power\_Grid}} &  \textsc{AstroPh} &  \textsc{BrightKite} &  \textsc{Flickr} \\
    \hline
     1 &        1.053 &    1.000 &       1.005 &   1.000 \\
    2  &        1.087 &    1.000 &       1.015 &   1.005 \\
    5  &        1.258 &    1.001 &       1.017 &   1.001 \\
    10 &        1.368 &    1.004 &       1.018 &   1.005 \\
    15 &        1.496 &    1.003 &       1.022 &   1.000 \\
    \hline
    \end{tabular}
\caption{Values of $\gamma$ as $k$ increases in the Uniform model with $p = 0.3$.} 
~\label{tbl:lcc-ratio-uniform-03}
\end{table}

The degree outperforms NetShield on the \textsc{AstroPh} and \textsc{BrightKite} datasets in the Uniform model with $p = 0.1$; on the other side, in the Uniform model with $p = 0.3$ and $p = 0.5$ $\gamma$ values around one have been reported; thus, the reduction in coverage due to the degree is almost equal to the one associated with Netshield.

\begin{table}
\centering
    \begin{tabular}{ccccc}
     \hline \hline 
    $k$ &  \textsc{{US\_Power\_Grid}} &  \textsc{AstroPh} &  \textsc{BrightKite} &  \textsc{Flickr} \\
    \hline
    1  &        1.040 &    1.000 &       1.004 &   1.000 \\
    2  &        1.036 &    1.000 &       1.005 &   1.004 \\
    5  &        1.136 &    1.000 &       1.007 &   1.000 \\
    10 &        1.573 &    1.001 &       1.012 &   1.005 \\
    15 &        1.613 &    1.003 &       1.009 &   1.000 \\
    \hline
    \end{tabular}
\caption{Values of $\gamma$ as $k$ increases in the Uniform model with $p = 0.5$.} 
~\label{tbl:lcc-ratio-uniform-05}
\end{table}

In the BC model, experimental results highlight an almost perfect alignment between the coverage of NetShield and the one of the degree, confirmed by $\gamma$ values close to one.

\begin{table}
\centering
    \begin{tabular}{ccccc}
     \hline \hline 
    $k$ &  \textsc{{US\_Power\_Grid}} &  \textsc{AstroPh} &  \textsc{BrightKite} &  \textsc{Flickr} \\
    \hline
  1 &        1.012 &      1.0 &         1.0 &     1.0 \\
  2 &        1.012 &      1.0 &         1.0 &     1.0 \\
  5 &        1.012 &      1.0 &         1.0 &     1.0 \\
 10 &        1.014 &      1.0 &         1.0 &     1.0 \\
 15 &        1.020 &      1.0 &         1.0 &     1.0 \\
   \hline
    \end{tabular}
\caption{Values of $\gamma$ as $k$ increases in the BC model.} 
~\label{tbl:lcc-ratio-best-connected}
\end{table}

\subsection{Summary of key results} %titolo provvisorio
\label{sub:takehome}
In short, the take-home message from the experiments herein conducted is as follows: 
\begin{enumerate*}[label=(\roman*)]
\item Degree is a centrality metric that, on average, produces the largest drop in both $\lambda_1$ and $c$. Degree is, in addition, a viable alternative to other methods (such as the NetShield algorithm) to detect group of nodes whose removal yields a relevant drop in the spectral radius.
\item The BC model guarantees a large drop in $\lambda_1$, even when only a small fraction of nodes actually fail.
\item In graphs deriving from human interactions and collaborations (such as \textsc{BrightKite}, \textsc{Flickr} and \textsc{AstroPh}) emerged a bigger drop in $\lambda_1$ than in the other datasets.
\item The degree, h-index, and coreness exhibit the same behaviour in social and collaborative networks, thus confirming insights provided in~\cite{lu2015hindex}.
\item The coverage analysis confirms that degree and coreness are better than the other centrality metrics herein considered, in terms of ability to fragment a graph into smaller and disjoint subcomponents, even when only a small fraction of nodes is targeted.
\end{enumerate*}
 In some datasets, a large gap in coverage reduction between degree and coreness centrality, and other centrality metrics has been appreciated; whereas in other datasets this gap appears to be more softened.

\section{Discussion}
\label{sec:implications}

This section illustrates the practical implications of our study.

It could be observed that the survival probability $p$ of a particular node in a graph $G$ can be interpreted as the {\em cost} to remove that node. More specifically, the higher the survival probability of a node, the higher the cost of its removal.

In the Uniform model, the cost to remove a node is the same across all nodes. Therefore, this study suggests that {\em the choice of targeting high degree nodes is always the best one}, independently of the value of the survival probability $p$. If $p$ were zero, the node removal task would be always successful; this case is well-known in the literature~\cite{lu2016vital} and the conclusions of previous studies are consistent with our findings: in real-world systems (e.g., the transportation system of a large city), large degree nodes (often called {\em hubs}) are the most important points of failure and, thus, an adequate protection of hubs leads to an effective protection of the whole system.

In the Best Connected (BC) model, the cost to remove a node is proportional to its degree. As an example, the cost to remove higher degree nodes in power-law graphs can be some orders of magnitude bigger than the cost to remove lower-degree nodes. Suppose now to have a budget $B$ to cover costs associated with the node removal task, which is insufficient to remove as many nodes as desired. Contrary to the Uniform model, the strategy of targeting high degree nodes might not be optimal, so a careful estimation of the trade-off between the costs of node removal and the corresponding loss in connectivity should be considered. For instance, Network Science methods have been widely applied to describe the structure of criminal organizations~\cite{agreste2016network,mastrobuoni2015value,villani2018virtuous}, and recent results indicate that high degree nodes in a criminal organization might not correspond to the major players in that organization. A repressive action which concentrates all the budget in removing high degree nodes might be ineffective because the nodes corresponding to the major players would not be under attack. Thus, the criminal organization is still alive and fully operative after high-degree node removal and, thus, the commitment of the whole budget $B$ to imprison high degree nodes could lead to a waste of time and financial resources.

Table~\ref{tbl:eval_banckmark} (resp. Table~\ref{tbl:cc_banckmark}) explicitly shows the percent deviation between the always successful node removal process (i.e., the Benchmark with $p=0$) and our probabilistic models in effectiveness (resp. coverage) experiments. These results confirmed how significant is the difference between a more realistic approach from the Benchmark one. Indeed, it varies a lot accordingly form the type of dataset, and the model considered reaching a gap that ranges from 2\% to 80\%.

\begin{table}
\centering
    \begin{tabular}{clrrl}
        \hline \hline 
       Dataset & $p=0.1$ & \ $p=0.3$ & \ $p=0.5$ & \ BC \\ 
        \hline 
        \textsc{US\_Power\_Grid} & 2.02\% & 26.54\% & 35.82\% & 35.69\%\\
        \textsc{AstroPh} & 2.17\% & 11.39\% & 32.16\% & 22.42\%\\
        \textsc{BrightKite} & 43.13\% & 66.57\% & 78.32\% & 62.26\%\\
        \textsc{Flickr} & 44.07\% & 70.07\% & 80.31\% & 49.47\%\\
    \hline
    \end{tabular}
\caption{Effectiveness comparison between probabilistic and classical approach.} 
~\label{tbl:eval_banckmark}
\end{table}

\begin{table}
\centering
    \begin{tabular}{clrrl}
        \hline \hline 
       Dataset & $p=0.1$ & \ $p=0.3$ & \ $p=0.5$ & \ BC \\ 
        \hline 
        \textsc{US\_Power\_Grid} & 24.82\% & 53.89\% & 65.88\% & 59.97\%\\
        \textsc{AstroPh} & 2.75\% & 6.16\% & 8.90\% & 4.28\%\\
        \textsc{BrightKite} & 29.39\% & 39.34\% & 44.49\% & 28.43\%\\
        \textsc{Flickr} & 4.88\% & 9.67\% & 12.31\% & 4.33\%\\
    \hline
    \end{tabular}
\caption{Coverage comparison between probabilistic and classical approach.} 
~\label{tbl:cc_banckmark}
\end{table}

%se%!TEX root = ./resilience_under_probabilistic_demeo.tex

\section{Conclusions and future works}
\label{sec:conclusions}

In this paper, a probabilistic model to describe node failure in graphs has been introduced, including two variants dubbed Uniform and Best Connected (BC).
Five popular centrality metrics have been considered (degree, h-index, coreness, Eigenvector, and Katz centrality), comparing their ability in reducing both the spectral radius $\lambda_1$ as well as the largest connected component size $c$.

The main finding has been that degree, but also coreness, can generally determine the biggest $\lambda_1$ drop, particularly in graphs deriving from human collaboration (e.g., \textsc{BrightKite} and \textsc{Flickr}). When it comes to social and collaborative networks, the h-index centrality is very effective too.

Furthermore, the experiments herein conducted unveiled an significant difference from the best metric, among our approaches (i.e., degree centrality), and the state-of-art strategy (i.e., always successful node removal through degree centrality). It confirms our hypothesis that the non-probabilistic approach is unrealistic because it does not take into account the cost of the node removal process. Thus, it leads to the unrealistic prevision of a faster network fragmentation.

In the next stages of this project, our plan is to extend this analysis to the variation of centrality metrics in connection to edge removal~\cite{GuQuHi19}. A challenge will be to find a suitable probabilistic model that would mirror the performance of the Uniform and the BC strategies described in this paper.
Indeed, in real Online Social Networks (OSNs) such as Facebook or Twitter, node failures reflect the deactivation of users' accounts, which is often a non-desirable effect. Yet, a range of other problems are linked to the interruptions in information flows, messages and status updates, which may be better captured by edge (rather than node) failures.

\bibliographystyle{unsrt}  


\end{document}